\documentclass[aps,prd,twocolumn,groupedaddress,nofootinbib]{revtex4-1}
\usepackage{amsmath}
\usepackage{amsfonts}
\usepackage{amssymb}
\usepackage{braket}
\usepackage{hyperref}

\newcommand{\I}{\mathrm{i}}
\newcommand{\e}{\mathrm{e}}
\newcommand{\D}{\mathrm{d}}
\newcommand{\op}[1]{\hat{#1}}
\renewcommand\vec\mathbf

\begin{document}

\title{Construction of quantum Dirac observables\\ and the emergence of WKB time}
\author{Leonardo Chataignier}
\email[]{lcmr@thp.uni-koeln.de}
\affiliation{Institute for Theoretical Physics, University of Cologne, Z\"{u}lpicher Stra\ss e 77, 50937 K\"{o}ln, Germany}

\date{\today}

\begin{abstract}
We describe a method of construction of gauge-invariant operators (Dirac observables or ``evolving constants of motion'') from the knowledge of the eigenstates of the gauge generator in time-reparametrization invariant mechanical systems. These invariant operators evolve unitarily with respect to an arbitrarily chosen time variable. We emphasize that the dynamics is relational, both in the classical and quantum theories. In this framework, we show how the ``emergent Wentzel-Kramers-Brillouin time'' often employed in quantum cosmology arises from a weak-coupling expansion of invariant transition amplitudes, and we illustrate an example of singularity avoidance in a vacuum Bianchi I (Kasner) model.
\end{abstract}
\maketitle

\section{Introduction}

One of the most important issues in any attempt to quantize the gravitational field is the proper understanding of the gauge symmetry of the theory (``general covariance'') at the quantum level. Presumably, a \emph{complete} quantum theory of gravity would involve a Hilbert space of physical states and a set of linear operators which would represent the observables of the theory. Both states and observables should transform appropriately under gauge transformations. There is no general agreement on how such a theory should be constructed or even if a Hilbert space is really necessary~\cite{Kiefer:book}, although there are multiple approaches being actively pursued~\cite{Oriti:book}.

Classically, the diffeomorphism symmetry induces a group of transformations in phase space (the Bergmann-Komar group~\cite{Bergmann:1972,Pons:2010}) and it is associated with a set of first class constraints~\cite{Rosenfeld:1930,Dirac:1950,Dirac:1958-1,Dirac:1958-2,Dirac:lectures,ADM:1962,Sundermeyer:1982,HT:book}. The gauge transformations (Lie derivatives) are generated in phase space by a combination of these constraints~\cite{Rosenfeld:1930,Pons:2010,Anderson:1951,Castellani:1982}. In this way, phase-space functions that are gauge invariant must have vanishing Lie derivatives and, if these functions do not depend explicitly on the spacetime coordinates, then they Poisson commute with the constraint functions. In this case, they are often called Dirac observables.

An important class of such observables is given by ``evolving constants of motion''~\cite{Rovelli:1991}, which are phase-space functions that encode the relational evolution between tensor fields according to the appropriate field equations (e.g., the Einstein field equations in general relativity). The evolving constants can be understood as gauge-invariant extensions of noninvariant quantities given in a particular frame~\cite{Woodard:1985,Teitelboim:1992,HT:book,Woodard:1993,Dittrich:2004,Dittrich:2005} and have been contemplated in the literature in the context of canonical gravity in both (quantum) geometrodynamics and loop quantum gravity~\cite{Carlip:1990,Gambini:2015-1,Gambini:2015-2}.

If one takes the view that the physical content of a generally covariant theory is entirely encoded in such relational phase-space functions, it is indeed reasonable to construct a canonical quantum theory based on operators that represent Dirac observables and physical states that are superpositions of eigenstates of the evolving constants. In this framework, the physical Hilbert space is the vector space of wave functions that are annihilated by the constraint operators. This apparently leads to a ``problem of time''~\cite{Kuchar:1991,Isham:1992}: physical states seem to be time independent and one has the impression that the dynamics is ``frozen.''

While there are many possible solutions to this problem (see, for instance,~\cite{Anderson:book} and references therein), it is arguably sufficient to note that the quantum dynamics has to be relational, as it is in the classical theory. The dynamics is not frozen, but rather encoded in the relational evolution of Dirac observables. Variations of this argument have been explored in the literature~\cite{Rovelli:1990-1,Rovelli:1990-2,Hartle:1996,Gambini:2001-1,Gambini:2001-2,Gambini:2009,Rovelli:2011,Tambornino:2012,Rovelli:2014,Hoehn:2018-1,Hoehn:2018-2,Hoehn:2019,Diaz:2019,Lusanna:book}, but a systematic way to construct Dirac observables is lacking. Moreover, the explicit connection between this relational view and other approaches to the ``problem of time'' has remained unclear. In particular, a popular ``solution'' is the ``semiclassical emergence of time'': time only exists when the wave function(al) of the gravitational field is in a semiclassical regime (see~\cite{Kuchar:1991,Isham:1992,Kiefer:1993-1,Kiefer:1993-2} for a review and~\cite{Halliwell:1984,Kiefer:2011-1,Brizuela:2015-1,Brizuela:2015-2,Kamenshchik:2013,Kamenshchik:2014,Kamenshchik:2016,Kamenshchik:2017} for phenomenological applications). For this reason, this emergent semiclassical time is often referred to as ``Wentzel-Kramers-Brillouin (WKB) time''~\cite{Zeh:1987,Kiefer:1993-1,Kiefer:1993-2}. Recently~\cite{Chataig:2019}, the author has argued that such a ``semiclassical approach to the problem of time'' coincides with a particular choice of gauge (i.e., time coordinate) and can be extended beyond the semiclassical level.

The purpose of this article is twofold: (1) to discuss a systematic, model-independent method of construction of gauge-invariant operators in covariant quantum mechanics, i.e., operators that commute with the gauge generator and therefore have a physical interpretation that is independent of the time parametrization (gauge) adopted; (2) to relate the heuristic notion of an ``emergent semiclassical time'' to the concrete and more fundamental framework in which the basic objects of the quantum theory are correlation functions of gauge-invariant operators.

The construction of operators corresponding to Dirac observables will be guided by an analogy to the Faddeev-Popov gauge-fixing method~\cite{FP-1,FP-2} in conventional gauge theories (i.e., Yang-Mills theories). Our approach will be canonical (operator-based) and we will not make use of path integrals. Although the restriction to mechanical theories is for the sake of simplicity, the method here presented is directly applicable to all minisuperspace models of quantum cosmology and, hence, it is useful. The field-theoretical case (with the possible issues of regularization and anomalies) will be left for future work.

The second objective is a continuation of~\cite{Chataig:2019}, in which it was extensively argued how the results of the usual semiclassical approach to the problem of time can be recovered from a complete quantum theory where the notion of ``gauge fixing'' was paramount. The work of~\cite{Chataig:2019} was, however, limited by the use of the indefinite Klein-Gordon inner product in the Hilbert space of physical states. In the present article, we adopt the positive-definite Rieffel induced inner product~\cite{Rieffel:1974,Landsman:1995,Marolf:1995-4,Marolf:1997,Marolf:2000,Halliwell:2001,Halliwell:2002}~(see also~\cite{Stueckelberg:1941-1,Stueckelberg:1942,Nambu:1950,Feynman:1950}), and we show how the emergence of WKB time occurs in the simple example of a relativistic particle, which is sufficient to illustrate the connection between the semiclassical time and the exact relational dynamics at the fully quantum level.

The article is structured in the following way. In Sec.~\ref{sec:general}, we review the classical theory and present the general formalism for the construction of Dirac observables in covariant quantum mechanics, comparing it with previous proposals. In the subsequent sections, we analyze concrete examples of this construction. In Sec.~\ref{sec:rel}, the relativistic particle is quantized and the corresponding Dirac observables are constructed. We show how they coincide with their nonrelativistic counterparts in the appropriate limit, in which the WKB time also emerges. In particular, we construct the ``time-of-arrival'' Dirac observable (see~\cite{Grot:1996,Hoehn:2018-2} and references therein) in the nonrelativistic limit. In Sec.~\ref{sec:kasner}, we analyze a cosmological model, the vacuum Bianchi I (Kasner) universe, and give an example of how the classical singularity may be avoided in the quantum theory. Finally, in Sec.~\ref{sec:conclusions}, we summarize our results and present our conclusions. We keep factors of $c$ and $\hbar$ explicit.

\section{\label{sec:general}The General Framework}
\subsection{Classical theory}
\subsubsection{\label{sec:observables-classical}Observables}
In preparation to the quantum theory of Dirac observables, we review the fundamentals of generally covariant classical mechanics, which can be regarded as a toy model of general relativity in $0+1$ spacetime dimensions~\cite{Teitelboim:1982}. We refer to the one-dimensional background manifold as the worldline. Given two worldline vectors $V_{(1,2)} = \epsilon_{(1,2)}(\tau)\frac{\D}{\D\tau}$, we can define the (intrinsic) metric on the worldline as $g(V_{(1)},V_{(2)}) = e^2(\tau)\epsilon_{(1)}(\tau)\epsilon_{(2)}(\tau)$, where $e(\tau)$ is a worldline scalar density called the einbein. Gauge transformations are worldline diffeomorphisms generated by a vector field $V = \epsilon(\tau)\frac{\D}{\D\tau}$~\cite{Bergmann:1972,Pons:2010}.

The dynamical variables are worldline tensors described in an arbitrary ``frame'' related to the choice of the worldline parameter $\tau$. Under reparametrizations of the worldline, the components of tensors transform covariantly. In fact, there is no problem in defining physical quantities (observables) to be covariant rather than invariant under worldline reparametrizations (see the discussions in~\cite{Pitts:2016,Pitts:2018,Pitts:2019} as well as in~\cite{Pons:2005,Pons:2009}). Thus, we can define observables to be worldline tensors. However, since one promotes the \emph{initial values} of dynamical fields to operators in the quantum theory, we would like to be able to describe the initial values independently of the choice of parametrization and, thus, in a gauge-invariant manner. In this way, we would like to construct Dirac observables, i.e., objects which commute with the phase-space constraints, to represent the \emph{invariant extensions} of initial values of worldline tensors. These extensions will then be promoted to operators in the quantum theory. Let us see how this can be achieved systematically.

For simplicity, we assume the fundamental dynamical fields are worldline scalars. The gauge transformation of a scalar field $q(\tau)$ reads
\begin{equation}\label{eq:general-scalar-transf}
\delta_{\epsilon(\tau)}q(\tau) :=\pounds_V q(\tau) = \epsilon(\tau)\frac{\D q(\tau)}{\D \tau} \ .
\end{equation}
For the dynamics to be reparametrization invariant, the Lagrangian $\mathcal{L}(q(\tau),\dot{q}(\tau))$ (where $\cdot \equiv \frac{\D}{\D \tau}$) must be a worldline scalar density, such that it transforms as follows~\cite{Teitelboim:1982,HT:book}:
\begin{equation}\label{eq:general-density-transf}
\delta_{\epsilon(\tau)}\mathcal{L} :=\pounds_V \mathcal{L} =  \frac{\D}{\D \tau}\left(\epsilon(\tau)\mathcal{L}\right) \ .
\end{equation}
This implies that the action
\begin{equation}
S = \int_a^b\D\tau\ \mathcal{L}(q(\tau),\dot{q}(\tau)) \ ,
\end{equation}
is invariant if the infinitesimal diffeomorphism $\epsilon(\tau)$ vanishes at the end points, i.e., $\epsilon(a) = \epsilon(b) = 0$. Otherwise, it is necessary to add boundary terms to the action to make it invariant~\cite{HT-Vergara:1992}. In fact, given a worldline one-form $\omega(\tau)\D\tau$, where $\omega(\tau)$ transforms as in~(\ref{eq:general-density-transf}), then the quantity
\begin{equation}\label{eq:general-observables}
\mathcal{O}_{\omega} = \int_{\alpha}^{\beta}\D\tau\ \omega(\tau)
\end{equation}
is an invariant (hence, observable) provided the integral converges and suitable boundary conditions are chosen for $\epsilon(\tau)$ and $\omega(\tau)$. For example, one may restrict $\epsilon(\tau)$ and $\omega(\tau)$ to periodic boundary conditions $\epsilon(\alpha) = \epsilon(\beta), \omega(\alpha) = \omega(\beta)$. Similarly, one may let $\alpha\to-\infty$ and $\beta\to+\infty$ if $\omega(\tau)$ is integrable and $\lim_{|\tau|\to\infty}\omega(\tau)\epsilon(\tau) = 0$. Objects of this form have been considered in~\cite{DeWitt:1962,Marolf:1995-1,Marolf:1995-2,Giddings:2006}.

An important class of observables is given by the evolving constants~\cite{Rovelli:1991,HT:book}, as mentioned in the Introduction. These objects encode the relational evolution of \emph{on-shell} tensor fields, i.e., fields which are solutions to the equations of motion, and they yield invariant extensions of the initial values. They can be constructed by imposing a gauge condition, i.e., by defining a parametrization of the worldline, in the following way. Let $\tau$ be an arbitrary initial parameter and define $s$ as a new time coordinate through the equation
\begin{equation}\label{eq:general-gauge-condition}
\chi\left(q(\tau),\dot{q}(\tau),e(\tau)\right) = s \ ,
\end{equation}
where $\chi$ is a worldline scalar that will be referred to as the gauge condition.\footnote{Gauge conditions of the form given in~(\ref{eq:general-gauge-condition}) are sufficient for our purposes, although more general gauge conditions are possible~\cite{HT:book}.} The condition is admissible if
\begin{equation}\label{eq:general-admissible-gauge}
\Delta_{\chi} := \frac{\D\chi}{\D\tau}\neq 0 \ ,
\end{equation}
which may be fulfilled only locally in the configuration-velocity space. In a region where~(\ref{eq:general-admissible-gauge}) holds, one can solve~(\ref{eq:general-gauge-condition}) for $\tau$ to find the coordinate transformation
\begin{equation}\label{eq:general-coordinate-transf}
\tau = \phi(q(0),\dot{q}(0),e(0),s) \ .
\end{equation}
If the gauge condition is admissible, $\phi$ defines a (field-dependent) diffeomorphism on the worldline, with which we can pull back tensor fields. The invariant extensions of initial values can then be obtained by writing the pullback in an arbitrary parametrization.\footnote{This corresponds to the statement that invariant extensions are obtained by writing gauge-fixed quantities in an arbitrary gauge.} To make this statement more precise, let us define the Dirac delta distribution
\begin{equation}
    \begin{aligned}
    &\delta(\tau) = 0 \  (\tau \neq 0) \ , \\
    \int_{-\infty}^{\infty}\mathrm{d}\tau \ &\delta(\tau)f(\tau) = f(0) \ .
    \end{aligned}
\end{equation}
Then, given a scalar field $f(\tau)$, we can write~[cf.~(\ref{eq:general-observables})]
\begin{equation}\label{eq:general-relational-scalar-classical}
    \begin{aligned}
    &\mathcal{O}[f|\chi = s]:= \phi^*f = \left.f(\tau)\right|_{\tau = \phi}\\ &= \int_{-\infty}^{\infty}\mathrm{d}\tau\ \delta(\tau-\phi(q(0),\dot{q}(0),e(0),s))f(\tau)\\
    &=\int_{-\infty}^{\infty}\mathrm{d}\tau\ \left|\frac{\mathrm{d}\chi}{\mathrm{d}\tau}\right|\delta(\chi(q(\tau),\dot{q}(\tau),e(\tau))-s)f(\tau) \ ,
    \end{aligned}
\end{equation}
provided~(\ref{eq:general-admissible-gauge}) holds and the integral in~(\ref{eq:general-relational-scalar-classical}) converges. Similar integral expressions have been considered in~\cite{DeWitt:1962,Marolf:1995-1,Marolf:1995-2,Giddings:2006}. For any \emph{fixed} value of $s = s_0$, Eq.~(\ref{eq:general-relational-scalar-classical}) defines an invariant extension (Dirac observable) of the initial value of $\phi^*f|_{s = s_0}$ in the sense that it is manifestly independent of the choice of $\tau$. We will see in Sec.~\ref{sec:evolving-invariant} that this property implies that the quantity given in~(\ref{eq:general-relational-scalar-classical}) Poisson commutes with the phase-space constraint. In particular, the Dirac observable associated with the identity function is again the identity
\begin{align}
    &\mathcal{O}[1|\chi = s] = \int_{-\infty}^{\infty}\mathrm{d}\tau\left|\frac{\mathrm{d}\chi}{\mathrm{d}\tau}\right|\delta(\chi(q(\tau),\dot{q}(\tau),e(\tau))-s)\notag\\
    &= \int_{-\infty}^{\infty}\mathrm{d}\tau\ \delta(\tau-\phi(q(0),\dot{q}(0),e(0),s)) = 1 \ . \label{eq:FP-classical}
\end{align}
Equation~(\ref{eq:FP-classical}) is the ``Faddeev-Popov resolution of the identity'' for the gauge condition $\chi$. The Dirac observable associated with the gauge condition itself is trivial
\begin{equation}\label{eq:DObs-chi-trivial}
\begin{aligned}
&\mathcal{O}[\chi|\chi = s] = \phi^*\chi = \int_{-\infty}^{\infty}\mathrm{d}\tau\left|\frac{\mathrm{d}\chi}{\mathrm{d}\tau}\right|\\&\times\delta(\chi(q(\tau),\dot{q}(\tau),e(\tau))-s)\chi(q(\tau),\dot{q}(\tau),e(\tau))\\
&= s \ .
\end{aligned}
\end{equation}
Similarly, given a one-form $\omega(\tau)\D\tau$, we can define the Dirac observable 
\begin{equation}\label{eq:general-relational-density-classical}
\begin{aligned}
    &\mathcal{O}[\omega|\chi = s] := \phi^*\omega\\ &= \int_{-\infty}^{\infty}\mathrm{d}\tau\ \left|\frac{\mathrm{d}\chi}{\mathrm{d}\tau}\right|\delta(\chi(q(\tau),\dot{q}(\tau),e(\tau))-s)\frac{\mathrm{d}\phi}{\mathrm{d}s}\omega(\tau) \ .
\end{aligned}
\end{equation}
As already noted, the integral expressions~(\ref{eq:general-relational-scalar-classical}) and~(\ref{eq:general-relational-density-classical}) are manifestly independent of the choice of $\tau$ and, thus, are gauge-invariant extensions for a fixed value of $s = s_0$. However, they generally depend on the gauge condition $\chi$ given in~(\ref{eq:general-gauge-condition}). This is usually the case with invariant extensions~\cite{Woodard:1985,Woodard:1993,HT:book, Teitelboim:1992}; i.e., they yield gauge-invariant but not gauge-independent objects. The physical interpretation of this procedure is particularly clear for the scalar Dirac observables [cf.~(\ref{eq:general-relational-scalar-classical})]: they represent the value of the scalar field $f$ ``when'' the scalar $\chi$ has the value $s_0$, i.e., they encode the (on-shell) relational evolution between the scalar fields.

The integral expressions~(\ref{eq:general-relational-scalar-classical}) and~(\ref{eq:general-relational-density-classical}) are most convenient for the quantization of Dirac observables that will be performed in Sec.~\ref{sec:general-quantum}. However, before quantizing the system we must analyze its dynamics in phase space.

\subsubsection{Hamiltonian and gauge generator}
If the fundamental fields $q(\tau)$ are worldline scalars, the Hamiltonian vanishes~\cite{Teitelboim:1982,HT:book}. To see this, we follow~\cite{Teitelboim:1982} and expand~(\ref{eq:general-density-transf}) and use~(\ref{eq:general-scalar-transf}) to obtain\footnote{Summation over repeated indices is implied.}
\begin{align*}
    \frac{\D}{\D\tau}\left(\epsilon(\tau)\mathcal{L}\right) &= \delta_{\epsilon(\tau)}\mathcal{L}= \frac{\partial \mathcal{L}}{\partial q^i}\delta_{\epsilon(\tau)}q^i(\tau)+\frac{\partial \mathcal{L}}{\partial\dot{q}^i}\delta_{\epsilon(\tau)}\dot{q}^i\\
    &= \frac{\partial \mathcal{L}}{\partial q^i}\epsilon(\tau)\dot{q}^i(\tau)+\frac{\partial \mathcal{L}}{\partial\dot{q}^i}\epsilon(\tau)\ddot{q}^i+\frac{\partial \mathcal{L}}{\partial\dot{q}^i}\dot{\epsilon}(\tau)\dot{q}^i\\
    &=\frac{\D}{\D \tau}(\epsilon(\tau)\mathcal{L})+\dot{\epsilon}(\tau)\left(\frac{\partial \mathcal{L}}{\partial\dot{q}^i}\dot{q}^i-\mathcal{L}\right) \ ,
\end{align*}
which yields
\begin{equation}\label{eq:general-Hamiltonian-vanishes}
    \mathcal{H}(q(\tau),p(\tau)) = p_i(\tau)\dot{q}^i(\tau)-\mathcal{L}(q(\tau),\dot{q}(\tau)) = 0 \ ,
\end{equation}
where the momenta are defined in the usual way, $p_i(\tau) = \frac{\partial \mathcal{L}}{\partial\dot{q}^i}$, and are worldline scalars. Equation~(\ref{eq:general-Hamiltonian-vanishes}) also implies that the Lagrangian is singular~\cite{Sundermeyer:1982},
\begin{equation}
\frac{\partial^2\mathcal{L}}{\partial\dot{q}^i\partial\dot{q}^j}\dot{q}^j = 0 \ ,
\end{equation}
i.e., that one cannot invert $p_i(\tau) = \frac{\partial \mathcal{L}}{\partial\dot{q}^i}$ to find the velocities as functions of coordinates and momenta. This entails that the momenta are not independent and are generally related by constraints $C(q,p) = 0$\footnote{Constraints of this type are called primary in the usual Rosenfeld-Dirac-Bergmann algorithm~\cite{Rosenfeld:1930,Dirac:1950,Anderson:1951,Salisbury:2017}.}. For simplicity, we assume there is only one constraint, which amounts to imposing that the only gauge symmetry of the theory is given by the worldline diffeomorphisms. Thus, the constraint algebra is automatically first class and Abelian.

The constraint $C(q,p) = 0$ defines a surface in phase space. In principle, this surface may be equivalently defined by different constraint functions, such as $C^2 = 0$ or, in general, $f(C)=0$ where $f(0) = 0$. One may also adopt redundant descriptions, such as $C = 0, C^2 = 0$, but we exclude this possibility for convenience. Are all these definitions equally valid in order to describe the dynamics of the system in phase space? The answer is no. As described in~\cite{HT:book} (see, in particular, Chapter 1), one must impose restrictions on the constraint functions known as regularity conditions. For the simple case at hand, the regularity conditions lead to the requirement that the constraint surface be coverable by open regions, on which the constraint function $C(q,p)$ (locally) satisfies
\begin{equation}\label{eq:regularity-HT}
\frac{\partial C}{\partial q^i}\D q^i +\frac{\partial C}{\partial p_i}\D p_i \neq 0 \  
\end{equation}
on the constraint surface. This condition implies that, after a canonical transformation, one may (locally) take $C(q,p) = p_1$, although this is not necessary in practice. If $C(q,p)$ satisfies~(\ref{eq:regularity-HT}) on the constraint surface, then $f(C)$ [with $f(0) = 0$] also satisfies this condition if the derivative of the function $f$ is such that $f'(0) \neq 0$. In what follows, we assume that the constraint surface is defined by a function $C(q,p)$ that satisfies~(\ref{eq:regularity-HT}) on the constraint surface.

We also make use of Dirac's weak equality sign $\approx$ to denote identities that hold only on the constraint surface~\cite{Dirac:1958-1}. Thus $\mathcal{H} \approx 0$, since the canonical Hamiltonian is well defined only if $C(q,p) = 0$ and we can extend it off the constraint surface in an arbitrary manner. Hence, there is no loss of generality if we write $\mathcal{H} = \lambda(\tau;q(\tau),p(\tau))C(q(\tau),p(\tau))$, where $\lambda$ is an arbitrary worldline scalar density. This is justified by Theorem 1.1 of~\cite{HT:book}. For the case of interest here, this can be shown as follows (see Appendix 1.A of~\cite{HT:book} for further details). Suppose, for convenience, that $C(q,p) = p_1$ [this is possible, after a canonical transformation, due to~(\ref{eq:regularity-HT})]. If $\mathcal{H}(\tau;q^1,q^j,p_1,p_j)$, $(j>1)$ is a worldline scalar density that extends the canonical Hamiltonian such that $\left.\mathcal{H}\right|_{p_1 = 0} = 0$, then we may write
\begin{align*}
\mathcal{H}(\tau;q,p) &= \int_0^1\D x\ \frac{\D}{\D x}\mathcal{H}(\tau;q^1, q^j,xp_1,p_j)\\
&= p_1\int_0^1\frac{\D x}{x}\ \frac{\D}{\D p_1}\mathcal{H}(\tau;q^1, q^j,xp_1,p_j)\\
&=:C(q,p)\lambda(\tau;q,p) \ ,
\end{align*}
where the last line is a definition of $\lambda$. From now on, we assume $C(q,p)$ has a general form (after inverting the canonical transformation, if necessary). As the choice of einbein $e(\tau;q(\tau),p(\tau))$ is also arbitrary, we may choose $e(\tau;q(\tau),p(\tau))=\lambda(\tau;q(\tau),p(\tau))$, to obtain
\begin{equation}\label{eq:primary-Hamiltonian}
\mathcal{H}(\tau; q,p) := e(\tau;q(\tau),p(\tau)) C(q(\tau),p(\tau)) \ .
\end{equation}
In this manner, the evolution in $\tau$ of a phase-space function $g(\tau; q(\tau),p(\tau))$ is given by
\begin{equation}
\frac{\D g}{\D\tau} = \frac{\partial g}{\partial\tau}+\{g, e C\} \approx \frac{\partial g}{\partial\tau}+e\{g, C\} \ ,
\end{equation}
where $\{\cdot,\cdot\}$ is the Poisson bracket
\begin{equation}\label{eq:Poisson-br}
\{ g , h \} = \frac{\partial g}{\partial q^i}\frac{\partial h}{\partial p_i}-\frac{\partial h}{\partial q^i}\frac{\partial g}{\partial p_i} \ .
\end{equation}
Can gauge transformations be represented as canonical transformations in phase space? For worldline scalars $f(q(\tau),p(\tau))$ with no explicit $\tau$ dependence, we have
\begin{equation}\notag
    \delta_{\epsilon(\tau)}f = \epsilon(\tau)\frac{\D f}{\D \tau} = \epsilon(\tau)\{f,e C\}\approx\{f,\epsilon(\tau)e C\} =: \{f,G\} \ .
\end{equation}
Thus, the reparametrizations of such worldline scalars are \emph{on-shell} canonical transformations generated by $G(\tau;q(\tau),p(\tau)) = \epsilon(\tau)e(\tau;q(\tau),p(\tau))C(q(\tau),p(\tau))$ (called the gauge generator). For our present purposes, this is all that is needed.\footnote{Even if one allows $\epsilon(\tau)$ to depend on the canonical variables $q(\tau),p(\tau)$, it is not possible to reach the gauge condition~(\ref{eq:general-gauge-condition}) by a canonical transformation, i.e., $\phi^*\chi = s$ is not a canonical transformation. The reason for this is clarified in~\cite{Pons:2009}, where Pons {\it et al.} note that the map that produces the invariant extensions is not invertible, since it projects all the points in a gauge orbit to the same image where the gauge condition is satisfied. Hence, this map cannot be canonical. Using a formalism which is different from (but equivalent to) the one presented here, they show that the invariant extension can be seen as a limit of a one-parameter family of canonical transformations.} However, it is worth mentioning that it is possible to extend the phase space to include the einbein and its conjugate momentum $(e,p_e)$ as a canonical pair subject to the constraint $p_e = 0$. In this way, one can describe the gauge variations of worldline scalars and one-forms as on-shell canonical transformations generated by $G = \xi C + \dot{\xi} p_e$, where $\xi = \epsilon(\tau)e(\tau;q(\tau),p(\tau))$~\cite{Pons:2005,Pons:2009,Pons:2010,Castellani:1982}. 

\subsubsection{\label{sec:evolving-invariant}Evolving constants are invariant extensions}
We stated in Sec.~\ref{sec:observables-classical} that the quantity $\mathcal{O}[f|\chi = s]$ given by the integral expression in~(\ref{eq:general-relational-scalar-classical}) represents an invariant extension of $\phi^*f$ for each fixed value of $s = s_0$ because it is manifestly independent of the choice of the initial arbitrary parametrization $\tau$. As is well-known, this statement can be substantiated by proving that $\mathcal{O}[f|\chi = s_0]$ is a Dirac observable, i.e., it Poisson commutes with the phase-space constraint and, therefore, with the gauge generator. To do this, we first note that the phase-space constraint generates evolution in ``proper time,'' defined as $\eta:=\int\D\tau\ e(\tau)$. Indeed, if $g$ is a phase-space function with no explicit time dependence, we obtain~[cf.~(\ref{eq:primary-Hamiltonian})]
\begin{equation}\label{eq:proper-time-evolution}
\{g,C\} \approx \frac{1}{e}\{g,\mathcal{H}\} = \frac{1}{e}\frac{\D g}{\D\tau} =: \frac{\D g}{\D\eta} \ .
\end{equation}
Thus, we can write~(\ref{eq:general-relational-scalar-classical}) in terms of proper time,
\begin{equation}\label{eq:scalar-proper-time-classical}
\begin{aligned}
&\mathcal{O}[f|\chi = s_0] = \int_{-\infty}^{\infty}\mathrm{d}\eta\ \left|\frac{\mathrm{d}\chi}{\mathrm{d}\eta}\right|\\&\times\delta(\chi(q(\eta),p(\eta))-s_0)f(q(\eta),p(\eta))\\
&\equiv \int_{-\infty}^{\infty}\mathrm{d}\eta\ \omega[f|\chi = s_0] \ ,
\end{aligned}
\end{equation}
where we assumed the scalars $f, \chi$ have no explicit time dependence. From~(\ref{eq:proper-time-evolution}) and~(\ref{eq:scalar-proper-time-classical}), we obtain
\begin{equation}
\{\mathcal{O}[f|\chi = s_0], C\} \approx \int_{-\infty}^{\infty}\mathrm{d}\eta\ \frac{\D}{\D\eta}\omega[f|\chi = s_0] = 0 \ .
\end{equation}
This result holds if
\begin{align*}
\lim_{|\eta|\to\infty}\left|\frac{\mathrm{d}\chi}{\mathrm{d}\eta}\right|\delta(\chi(q(\eta),p(\eta))-s_0)f(q(\eta),p(\eta)) = 0
\end{align*}
for fixed values of $s_0$ and the initial conditions $q(0),p(0)$.

Since $\mathcal{O}[f|\chi = s]$ is a Dirac observable for each fixed value of $s$, one sees that there is a one-parameter family of invariant functions, each corresponding to one moment of the evolution. That is why such objects are often called evolving constants~\cite{Rovelli:1991}.

\subsubsection{\label{sec:Dynamics-Dobs}Dynamics of Dirac observables}
The pullback of on-shell scalar functions $f(q(\tau),p(\tau))$ under $\phi$ given in~(\ref{eq:general-coordinate-transf}) is evidently dynamical (time-dependent) in general. Indeed, let us write $\phi(q(0),\dot{q}(0),e(0),s)\equiv\phi(s)$ for brevity. We can then write
\begin{equation}\label{eq:evolution-classical}
\begin{aligned}
\frac{\D}{\D s}\mathcal{O}[f|\chi = s] &= \frac{\D}{\D s}f(q(\phi(s)),p(\phi(s)))\\
& = \frac{\D\phi}{\D s}\left[\frac{\D f}{\D\tau}\right]_{\tau = \phi}\\
& = \frac{\D\phi}{\D s}\left\{f,\mathcal{H}\right\}_{\tau = \phi} \ ,
\end{aligned}
\end{equation}
where both the Hamiltonian $\mathcal{H}$ [cf.~(\ref{eq:primary-Hamiltonian})] and the Poisson bracket [cf.~(\ref{eq:Poisson-br})] are taken with respect to the original set of fields $q(\tau)$ and $p(\tau)$ as opposed to the pulled-back fields. Only at the end of the calculation does one set $\tau = \phi(s)$ [and $C(q,p) = 0$]. Moreover, by setting $f = \chi$ in the above equation and using~(\ref{eq:DObs-chi-trivial}), we find
\begin{equation}\label{eq:gauge-fixed-lapse}
\frac{\D\phi}{\D s} = \frac{1}{\left\{\chi,\mathcal{H}\right\}_{\tau = \phi}} \ ,
\end{equation}
where $\left\{\chi,\mathcal{H}\right\}_{\tau = \phi}\neq0$ due to~(\ref{eq:general-admissible-gauge}). In the context of minisuperspace quantum cosmology, Eq.~(\ref{eq:gauge-fixed-lapse}) yields the gauge-fixed lapse function. If we insert~(\ref{eq:gauge-fixed-lapse}) on~(\ref{eq:evolution-classical}), we obtain
\begin{equation}\label{eq:gauge-fixed-eom}
\frac{\D}{\D s}\mathcal{O}[f|\chi = s] = \frac{1}{\left\{\chi,\mathcal{H}\right\}_{\tau = \phi}}\left\{f,\mathcal{H}\right\}_{\tau = \phi} \ ,
\end{equation}
which shows that the dynamics of observables is not frozen in general. In fact, Eq.~(\ref{eq:gauge-fixed-eom}) yields the gauge-fixed (or ``reduced'') equations of motion for the dynamical variables. The on-shell gauge-fixed evolution is generated by $\mathcal{H}_{gf}:=\frac{1}{\left\{\chi,\mathcal{H}\right\}}\mathcal{H}$. Indeed,
\begin{align*}
&\left\{\chi,\mathcal{H}_{gf}\right\} = \left\{\chi, \frac{1}{\left\{\chi,\mathcal{H}\right\}}\mathcal{H}\right\} \approx \frac{1}{\left\{\chi,\mathcal{H}\right\}}\left\{\chi,\mathcal{H}\right\} = 1 \ , \\
&\frac{\D}{\D s}\mathcal{O}[f|\chi = s] \approx \left\{f,\mathcal{H}_{gf}\right\}_{\tau = \phi} \ .
\end{align*}
Moreover, the right-hand side of~(\ref{eq:gauge-fixed-eom}) is a Dirac observable for each fixed value of $s$. To see this, we use the right-hand side of~(\ref{eq:general-relational-scalar-classical}) to write [cf.~(\ref{eq:general-relational-density-classical})]
\begin{align*}
&\frac{\D}{\D s}\mathcal{O}[f|\chi = s]\\& = \int_{-\infty}^{\infty}\mathrm{d}\tau\ \frac{\D}{\D s}\delta(\tau-\phi(q(0),\dot{q}(0),e(0),s))f(q(\tau),p(\tau))\\
&= \int_{-\infty}^{\infty}\mathrm{d}\tau\ \delta(\tau-\phi(q(0),\dot{q}(0),e(0),s))\frac{\D\phi}{\D s}\frac{\D f}{\D\tau}\\
&= \int_{-\infty}^{\infty}\mathrm{d}\tau\ \left|\frac{\mathrm{d}\chi}{\mathrm{d}\tau}\right|\delta(\chi(q(\tau),\dot{q}(\tau),e(\tau))-s)\frac{\D\phi}{\D s}\{f,\mathcal{H}\}\\
&\approx \mathcal{O}\left[\{f,\mathcal{H}_{gf}\}|\chi = s\right] \ .
\end{align*}
The equation
\begin{equation}\label{eq:gauge-fixed-eom-invariant}
\frac{\D}{\D s}\mathcal{O}[f|\chi = s] \approx \mathcal{O}\left[\{f,\mathcal{H}_{gf}\}|\chi = s\right] 
\end{equation}
was also obtained in~\cite{Pons:2009} using a different method. Equation~(\ref{eq:gauge-fixed-eom-invariant}) is of key importance for the quantum theory, since we expect that it can be promoted to a Heisenberg-picture equation of motion, both sides of which are well-defined operators (which commute with the constraint operator for each value of $s$). We will see in Secs.~\ref{sec:quantum-Dobs},~\ref{sec:rel-qDobs-1}, and~\ref{sec:rel-qDobs-2} how this can be achieved.

\subsection{\label{sec:general-quantum}Quantum theory}
\subsubsection{The physical Hilbert space}
Following~\cite{Rieffel:1974,Landsman:1995,Marolf:1995-4,Marolf:1997,Marolf:2000,Halliwell:2001,Halliwell:2002}, we promote the classical phase-space constraint $C(q,p)$ to a linear operator $\op{C}$ and assume that it is possible to choose the factor ordering such that $\op{C}$ is self-adjoint in an auxiliary Hilbert space of square-integrable functions equipped with an auxiliary inner product $\braket{\cdot|\cdot}$. In this way, $\op{C}$ has a complete orthonormal system of eigenstates
\begin{align}
\op{C}\ket{E,{\bf k}} &= E\ket{E,{\bf k}} \ , \\
\braket{E',\mathbf{k}'|E,\mathbf{k}} &= \delta(E', E)\delta(\mathbf{k}',\mathbf{k}) \ ,
\end{align}
where $\mathbf{k}$ labels degeneracies. The symbol $\delta(\cdot,\cdot)$ stands for a Kronecker or Dirac delta, depending on whether the spectrum of $\op{C}$ is discrete or continuous.

The quantum analogue of the classical constraint surface $C(q,p) = 0$ is the linear subspace of states in the kernel of $\op{C}$, which can be written as superpositions of $\ket{E = 0,{\bf k}}$. These states are invariant under the unitary flow of the constraint operator $\e^{\frac{\I}{\hbar}\tau\op{C}}\ket{E = 0,{\bf k}} = \ket{E = 0,{\bf k}}$ and their overlap reads 
\begin{equation}
\braket{E = 0, \mathbf{k}'|E = 0, \mathbf{k}} = \delta(0,0)\delta(\mathbf{k}',\mathbf{k}) \ .
\end{equation}
The factor of $\delta(0,0)$ is divergent if zero is in the continuous part of the spectrum of $\op{C}$, which implies the auxiliary inner product cannot be used in this subspace. It is possible~\cite{Rieffel:1974,Landsman:1995,Marolf:1995-4,Marolf:1997,Marolf:2000,Halliwell:2001,Halliwell:2002,HT-SUSY:1982} to define a regularized (induced) inner product $(\cdot|\cdot)$ on the kernel of $\op{C}$ in the following way:
\begin{equation}
\braket{E',\mathbf{k}'|E,\mathbf{k}} =: \delta(E', E)(E',\mathbf{k}'|E,\mathbf{k}) \ ,
\end{equation}
such that
\begin{equation}\label{eq:physical-inner-product}
(E = 0, \mathbf{k}'|E = 0, \mathbf{k}) = \delta(\mathbf{k}',\mathbf{k}) \ .
\end{equation}
Now consider the superpositions
\begin{equation}
\ket{\phi_E^{(1,2)}} = \sum_{\mathbf{k}}\phi^{(1,2)}(\mathbf{k})\ket{E,{\bf k}} \ , 
\end{equation}
where the sum over $\mathbf{k}$ must be replaced by an integral if the degeneracies are labeled by continuous indices. Then, from~(\ref{eq:physical-inner-product}), we obtain the (Rieffel induced) inner product for general invariant states
\begin{equation}\label{eq:physical-inner-product-2}
(\phi^{(1)}_{E = 0}|\phi^{(2)}_{E = 0}) = \sum_{\mathbf{k}}\bar{\phi}^{(1)}(\mathbf{k})\phi^{(2)}(\mathbf{k}) \ .
\end{equation}
The kernel of $\op{C}$ equipped with the inner product~(\ref{eq:physical-inner-product-2}) is defined to be the physical Hilbert space of the theory.

\subsubsection{\label{sec:quantum-Dobs}Matrix elements of quantum Dirac observables}
We are now in a position to propose a method of construction of operators that correspond to the classical Dirac observables. To begin with, according to~(\ref{eq:primary-Hamiltonian}), the Hamiltonian operator may be defined as $\op{\mathcal{H}} = e(\tau)\op{C}$, where factor ordering issues are avoided by choosing the arbitrary $\tau$-parametrization such that the einbein $e(\tau)$ is not a function of the canonical variables, and thus, it is a c-number in the quantum theory. The simplest choice is $e(\tau) = 1$ (proper time gauge), which is the one we adopt. In this way, any classical observable of the form given in~(\ref{eq:general-observables}) can be promoted to the operator
\begin{equation}
\op{\mathcal{O}}_{\omega} = \int_{\alpha}^{\beta}\mathrm{d}\tau\ \e^{\frac{\I}{\hbar}\tau\op{C}}\op{\omega}\ \e^{-\frac{\I}{\hbar}\tau\op{C}} \ .
\end{equation}
If the spectrum of $\op{C}$ is discrete, one may choose $\alpha = 0, \beta = 2\pi$, whereas if the spectrum is continuous, we let $\alpha\to-\infty$ and $\beta\to+\infty.$ In any case, we find the matrix elements
\begin{equation}\notag
\begin{aligned}
&\left<\phi_{E'}^{(1)}\left|\op{\mathcal{O}}_{\omega}\right|\phi_{E}^{(2)}\right>\\ &= 2\pi\hbar\delta(E',E)\sum_{\mathbf{k}',\mathbf{k}}\bar{\phi}^{(1)}(\mathbf{k}')\braket{E',\mathbf{k}'|\op{\omega}|E,\mathbf{k}}\phi^{(2)}(\mathbf{k}) \ ,
\end{aligned}
\end{equation}
and the regularized matrix elements are~\cite{Marolf:1995-1}
\begin{equation}\label{eq:physical-matrix-elements-DObs}
\begin{aligned}
&\frac{1}{2\pi\hbar}\left(\phi_{E = 0}^{(1)}\left|\op{\mathcal{O}}_{\omega}\right|\phi_{E = 0}^{(2)}\right)\\ &:= \sum_{\mathbf{k}',\mathbf{k}}\bar{\phi}^{(1)}(\mathbf{k}')\braket{E=0,\mathbf{k}'|\op{\omega}|E=0,\mathbf{k}}\phi^{(2)}(\mathbf{k}) \ .
\end{aligned}
\end{equation}
Now we would like to define an operator $\op{\omega}[f|\chi = s]$, such that\footnote{We assume that the spectrum of $\op{C}$ is continuous in what follows, since this will be the case in the concrete examples analyzed in later sections.}
\begin{equation}\label{eq:tentative-quantum-Dobs}
\op{\mathcal{O}}[f|\chi = s] = \int_{-\infty}^{\infty}\D\tau\ \e^{\frac{\I}{\hbar}\tau\op{C}}\op{\omega}[f|\chi = s]\e^{-\frac{\I}{\hbar}\tau\op{C}} 
\end{equation}
is a symmetric quantization of the classical scalar Dirac observable given in~(\ref{eq:general-relational-scalar-classical}). In particular, we require that an operator version of the Faddeev-Popov resolution of the identity~(\ref{eq:FP-classical}) holds, i.e.,
\begin{equation}\label{eq:FP-quantum-0}
\left(\phi_{E = 0}^{(1)}\left|\op{\mathcal{O}}[1|\chi = s]\right|\phi_{E = 0}^{(2)}\right) = \sum_{\mathbf{k}}\bar{\phi}^{(1)}(\mathbf{k})\phi^{(2)}(\mathbf{k}) \ ,
\end{equation}
which implies that the operator $\op{\omega}[1|\chi = s]$ must satisfy the relation
\begin{equation}\label{eq:FP-quantum}
 2\pi\hbar\braket{E=0,\mathbf{k}'|\op{\omega}[1|\chi = s]|E=0,\mathbf{k}} = \delta({\bf k}',{\bf k})
\end{equation}
for all values of $s$. In this point we differ from the work of Marolf in~\cite{Marolf:1995-1}, in which the definition of the operator given in~(\ref{eq:tentative-quantum-Dobs}) was chosen in such a way that the invariant extension of the identity was not the identity operator, a result which we consider to be undesirable. Indeed, for the case of the relativistic particle (which we will analyze in Sec.~\ref{sec:rel}), the operator definition chosen in~\cite{Marolf:1995-1} yields $\op{\mathcal{O}}[1|q^0 = c s] = \mathrm{sgn}(\op{p_0}) \neq \op{1}$. We believe that~(\ref{eq:FP-quantum}) should be the correct requirement. In fact, Eq.~(\ref{eq:FP-quantum}) is equivalent to regularizing the inner product by the ``insertion of an operator gauge condition,'' a procedure that was advocated in~\cite{HT:book,Woodard:1993}.

How can we define $\op{\omega}[1|\chi = s]$? Given a gauge condition operator $\op{\chi}$ which is self-adjoint in the auxiliary inner product, the (improper) projectors onto its eigenspaces are $\op{P}_{\chi = s} = \sum_{\bf n}\ket{\chi = s,\bf n}\bra{\chi = s,\bf n}$, where ${\bf n}$ labels degeneracies of the eigenstates of $\op{\chi}$. Since the classical gauge condition is admissible only if~(\ref{eq:general-admissible-gauge}) holds, i.e., if $\Delta_{\chi} = \{\chi, C\}\neq0$, we consider the operator
\begin{equation}
\sum_{\sigma = \pm}\Theta(\sigma\op{\Delta}_{\chi})\op{P}_{\chi = s}\Theta(\sigma\op{\Delta}_{\chi}) \ ,
\end{equation}
where the operators $\Theta(\sigma\op{\Delta}_{\chi})$ are included to project out the zero modes of $\op{\Delta}_{\chi}:=-\frac{\I}{\hbar}[\op{\chi},\op{C}]$. We can now define
\begin{equation}\label{eq:FP-quantum-inverse-measure}
\begin{aligned}
&\op{\mathcal{O}}[\left|\Delta_{\chi}\right|^{-1}|\chi = s]\\
&:= \sum_{\sigma = \pm}\int_{-\infty}^{\infty}\D\tau\ \e^{\frac{\I}{\hbar}\tau\op{C}}\Theta(\sigma\op{\Delta}_{\chi})\op{P}_{\chi = s}\Theta(\sigma\op{\Delta}_{\chi})\e^{-\frac{\I}{\hbar}\tau\op{C}}\\
&\equiv \left|\op{\Delta}_{\chi}^{\mathcal{O}}\right|^{-1} \ ,
\end{aligned}
\end{equation}
where we introduced the notation $\op{\Delta}_{\chi}^{\mathcal{O}}$ for brevity. The operator given in~(\ref{eq:FP-quantum-inverse-measure}) is by construction an invariant for each fixed value of $s$. Moreover, it is a symmetric quantization of the classical expression for the invariant extension of $\left|\Delta_{\chi}\right|^{-1}$, as can easily be verified. From~(\ref{eq:FP-quantum-inverse-measure}), we obtain the symmetric resolution of the identity
\begin{equation}\notag
\begin{aligned}
&\op{1} = \sum_{\sigma = \pm}\int_{-\infty}^{\infty}\D\tau\ \e^{\frac{\I}{\hbar}\tau\op{C}}\left|\op{\Delta}_{\chi}^{\mathcal{O}}\right|^{\frac{1}{2}}\Theta(\sigma\op{\Delta}_{\chi})\\
&\times\op{P}_{\chi = s}\Theta(\sigma\op{\Delta}_{\chi})\left|\op{\Delta}_{\chi}^{\mathcal{O}}\right|^{\frac{1}{2}}\e^{-\frac{\I}{\hbar}\tau\op{C}} =: \op{\mathcal{O}}[1|\chi = s] \ ,
\end{aligned}
\end{equation}
which leads to the sought-after definition
\begin{equation}\label{eq:FP-quantum-measure}
\begin{aligned}
&\op{\omega}[1|\chi = s]\\ &:= \sum_{\sigma = \pm}\left|\op{\Delta}_{\chi}^{\mathcal{O}}\right|^{\frac{1}{2}}\Theta(\sigma\op{\Delta}_{\chi})\op{P}_{\chi = s}\Theta(\sigma\op{\Delta}_{\chi})\left|\op{\Delta}_{\chi}^{\mathcal{O}}\right|^{\frac{1}{2}} \ .
\end{aligned}
\end{equation}
Equation~(\ref{eq:FP-quantum-measure}) is the canonical (operator-based) analogue of the usual Faddeev-Popov procedure employed in path integrals~\cite{FP-1,FP-2}. A similar canonical procedure was suggested in~\cite{Woodard:1993}, although it was not specified which factor ordering was required and the need to include the $\Theta(\sigma\op{\Delta}_{\chi})$ operators was not recognized.

We are now in a position to define invariant extensions of operators other than the identity. Suppose $\op{f}$ is a scalar operator which commutes with the gauge condition and is self-adjoint with respect to the auxiliary inner product. Then $\op{f}$ and $\op{\chi}$ share a complete orthonormal system of eigenstates $\ket{\chi,f,{\bf n}}$, where ${\bf n}$ labels other possible degeneracies. We may write
\begin{equation}
\op{P}_{\chi = s} = \sum_{f,{\bf n}}\ket{\chi = s, f,{\bf n}}\bra{\chi = s, f,{\bf n}} \ .
\end{equation}
In analogy to~(\ref{eq:FP-quantum-measure}), we \emph{define}
\begin{equation}\label{eq:general-relational-scalar-quantum}
\begin{aligned}
&\op{\omega}[f|\chi = s] := \sum_{\sigma = \pm, f, {\bf n}}f\ \left|\op{\Delta}_{\chi}^{\mathcal{O}}\right|^{\frac{1}{2}}\Theta(\sigma\op{\Delta}_{\chi})\\
&\times\ket{\chi = s, f,{\bf n}}\bra{\chi = s, f,{\bf n}}\Theta(\sigma\op{\Delta}_{\chi})\left|\op{\Delta}_{\chi}^{\mathcal{O}}\right|^{\frac{1}{2}} \ ,
\end{aligned}
\end{equation}
which amounts to defining the quantum Dirac observable $\op{\mathcal{O}}[f|\chi = s]$ [cf.~(\ref{eq:tentative-quantum-Dobs})] via its spectral decomposition. We can also define invariant extensions of scalars which do not commute with the gauge condition in the following way. We first note that the (improper) projector onto the eigenspaces of $\op{\chi}$ can be written as
\begin{equation}\label{eq:gauge-condition-projector-integral}
\begin{aligned}
\op{P}_{\chi = s} &= \sum_{\bf n}\ket{\chi = s,\bf n}\bra{\chi = s,\bf n}\\
& =  \sum_{\chi, \bf n}\delta(\chi,s)\ket{\chi,\bf n}\bra{\chi,\bf n}\\
&= \sum_{\chi, \bf n}\int_{\alpha}^{\beta}\frac{\D\lambda}{2\pi\hbar}\ \e^{\frac{\I}{\hbar}\lambda(\chi-s)}\ket{\chi,\bf n}\bra{\chi,\bf n}\\
&=\int_{\alpha}^{\beta}\frac{\D\lambda}{2\pi\hbar}\ \e^{\frac{\I}{2\hbar}\lambda(\op{\chi}-s\op{1})}\ \op{1}\ \e^{\frac{\I}{2\hbar}\lambda(\op{\chi}-s\op{1})} \ ,
\end{aligned}
\end{equation}
where $\alpha = 0, \beta = 2\pi$ if the spectrum of $\op{\chi}$ is discrete and $\alpha\to-\infty, \beta\to+\infty$ if the spectrum of $\op{\chi}$ is continuous. If $\op{f}$ is a self-adjoint scalar operator that does not commute with $\op{\chi}$, we can use~(\ref{eq:gauge-condition-projector-integral}) to generalize~(\ref{eq:general-relational-scalar-quantum}) to
\begin{equation}\label{eq:general-relational-scalar-quantum-2}
\begin{aligned}
&\op{\omega}[f|\chi = s] :=\sum_{\sigma = \pm}\int_{\alpha}^{\beta}\frac{\D\lambda}{2\pi\hbar}\ \left|\op{\Delta}_{\chi}^{\mathcal{O}}\right|^{\frac{1}{2}}\Theta(\sigma\op{\Delta}_{\chi})\\
&\times\e^{\frac{\I}{2\hbar}\lambda(\op{\chi}-s\op{1})}\ \op{f}\ \e^{\frac{\I}{2\hbar}\lambda(\op{\chi}-s\op{1})}\Theta(\sigma\op{\Delta}_{\chi})\left|\op{\Delta}_{\chi}^{\mathcal{O}}\right|^{\frac{1}{2}} \ .
\end{aligned}
\end{equation}
We take~(\ref{eq:tentative-quantum-Dobs}) together with~(\ref{eq:general-relational-scalar-quantum-2}) to be the general definition of the invariant extension of the scalar operator $\op{f}$. Invariant extensions of scalar densities could be defined in a similar way but we shall have no need for them in what follows. In the next sections, we will apply the general formalism here presented to the concrete examples of the relativistic free particle and the vacuum Bianchi I model.

\section{\label{sec:rel}The Relativistic Particle}
Let us now illustrate the general ideas presented in the previous section for the relativistic free particle, which is the archetypical example of a time-reparametrization invariant system. We first present the construction of Dirac observables in the classical theory and their nonrelativistic limit. We then quantize the theory and show that the notion of WKB time emerges in the nonrelativistic limit of invariant transition amplitudes. This result, although expected, clarifies the relation between the semiclassical approach to the problem of time and the more complete quantum theory based on the induced inner product.\footnote{The semiclassical approach was thoroughly analyzed in~\cite{Chataig:2019}. The example of the relativistic particle here presented serves to elucidate how this approach is related to the fundamental reparametrization invariance of the theory at the quantum level.} Moreover, we discuss the dynamics of quantum Dirac observables also in the nonrelativistic limit to compare the formalism here presented with the results of~\cite{Hoehn:2018-2,Hoehn:2019}.

\subsection{\label{sec:rel-classical}Classical theory}
\subsubsection{Observables}
The action for a massive relativistic particle moving in the $(d+1)$-dimensional Minkowski spacetime reads
\begin{equation}\label{eq:rel-action}
I = -m c\int_a^b \D\tau\ \sqrt{-\eta_{\mu\nu}\frac{\D q^{\mu}}{\D\tau}\frac{\D q^{\nu}}{\D\tau}} \ ,
\end{equation}
where $\eta_{\mu\nu} \ (\mu,\nu = 0,\ldots, d)$ are the coefficients of the Minkowski metric with signature $(-,+\cdots+)$ and $q^{\mu} = (c t, \vec{q})$ are the spacetime coordinates. The action is invariant under reparametrizations of $\tau$ which coincide with the identity at the end points. The Euler-Lagrange equations yield
\begin{equation}\label{eq:rel-velo}
\dot{q}^{\mu} = \eta^{\mu\nu}p_{\nu}\sqrt{-\eta_{\rho\lambda}\dot{q}^{\rho}\dot{q}^{\lambda}} \ ,
\end{equation}
where $p_{\mu} = \left(\frac{p_t}{c},\vec{p}\right)$ are constants that satisfy the initial-value constraint
\begin{equation}\label{eq:rel-constraint}
C = -\frac{p_t^2}{2c^2}+\frac{\vec{p}^2}{2}+\frac{m^2c^2}{2} = 0 \ .
\end{equation}
The solutions of~(\ref{eq:rel-velo}) are relational: we can determine the trajectories of one coordinate in terms of another. For example, we find
\begin{equation}\label{eq:rel-sol-1}
\vec{q}(\tau) =  \vec{q}(a)-\frac{c^2\vec{p}}{p_t}(t(\tau)-t(a))\ \  \ \ (p_t \neq 0) \ 
\end{equation}
by dividing the equation for $\dot{\vec{q}}$ by the equation for $\dot{q}^0 = c\dot{t}$. We note that the relation~(\ref{eq:rel-sol-1}) holds in any parametrization $\tau$ (any gauge). In this way, the boundary values $\vec{q}(a)$ may be seen as an invariant extension of $\vec{q}(\tau)$ for a fixed value of $t(a)$; i.e., 
\begin{equation}\label{eq:rel-DObs-1}
\vec{q}(a) = \vec{q}(\tau)+\frac{c^2\vec{p}}{p_t}(t(\tau)-t(a))
\end{equation}
is a Dirac observable. Indeed,
\begin{align*}
&\delta_{\epsilon(\tau)}\left[\vec{q}(\tau)+\frac{c^2\vec{p}}{p_t}(t(\tau)-t(a))\right]\\
& = \delta_{\epsilon(\tau)}\vec{q}(\tau)+\frac{c^2\vec{p}}{p_t}\delta_{\epsilon(\tau)}t(\tau)\\
&=\epsilon(\tau)\sqrt{-\eta_{\rho\lambda}\dot{q}^{\rho}\dot{q}^{\lambda}}\left[\vec{p}-\frac{c^2\vec{p}}{p_t}\frac{p_t}{c^2}\right] = 0 \ ,
\end{align*}
where we used~(\ref{eq:rel-velo}). The observable given in~(\ref{eq:rel-DObs-1}) represents the value of $\vec{q}(\tau)$ ``when'' $t(\tau) = t(a)$. This is true independently of the chosen parametrization $\tau$; i.e., it is a gauge-invariant statement. Similarly, we may construct an invariant extension of $t(\tau)$ by writing it in terms of $q^1(\tau)$. We obtain
\begin{equation}\label{eq:rel-DObs-2}
\begin{aligned}
t(a) &= t(\tau)+\frac{p_t}{c^2p_1}(q^1(\tau)-q^1(a)) \ , \\
q^j(a) &= q^j(\tau)-\frac{p_j}{p_1}(q^1(\tau)-q^1(a)) \ , \ (j = 2,...,d) \ .
\end{aligned}
\end{equation}
It is clear that the quantities given in~(\ref{eq:rel-DObs-2}) are invariants. The right-hand sides of~(\ref{eq:rel-DObs-2}) are well-defined provided $p_1\neq 0$. It is useful to note that the dynamics of~(\ref{eq:rel-DObs-1}) and~(\ref{eq:rel-DObs-2}) may be expressed in terms of Poisson brackets (as defined in~(\ref{eq:Poisson-br}) with the basic variables $q^{\mu}(\tau), p_{\mu}(\tau)$) in the following way:
\begin{align}
\frac{\partial\vec{q}(a)}{\partial t(a)} &= -\frac{c^2\vec{p}}{p_t} = \{p_t, \vec{q}(a)\} \ , \label{eq:PB-classical-1}\\
\frac{\partial t(a)}{\partial q^1(a)} &= -\frac{p_t}{c^2p_1} = \{p_1,t(a)\} \ , \label{eq:PB-classical-2}
\end{align}
and similarly for $q^j(a) \ (j = 2,\ldots,d)$. In Secs.~\ref{sec:rel-qDobs-1} and~\ref{sec:rel-qDobs-2}, we will find the quantum analogues of~(\ref{eq:PB-classical-1}) and~(\ref{eq:PB-classical-2}) as equations that determine the dynamics of quantum Dirac observables.

Finally, we may express~(\ref{eq:rel-DObs-1}) and~(\ref{eq:rel-DObs-2}) in integral form, which will be useful in the quantum theory~(cf. Sec.~\ref{sec:general-quantum}). For example,
\begin{equation}\label{eq:rel-DObs-int}
\begin{aligned}
\vec{q}(a) &= \vec{q}(\tau)|_{\tau = a} = \int_{-\infty}^{\infty}\D\tau\ \delta(\tau-a)\vec{q}(\tau)\\
&=\int_{-\infty}^{\infty}\D\tau\ \left|\frac{\D t}{\D\tau}\right| \delta(t(\tau)-t(a))\vec{q}(\tau) \ ,
\end{aligned}
\end{equation}
and similarly for~(\ref{eq:rel-DObs-2}). Evidently, Eq.~(\ref{eq:rel-DObs-int}) holds only if $\left|\frac{\D t}{\D\tau}\right|\neq0$, i.e., if the gauge condition $t(\tau) = t(a)$ is well-defined.

\subsubsection{On-shell action and the Hamilton-Jacobi constraint}
The constants $p_{\mu}$ may be eliminated in terms of the boundary values $q^{\mu}(a), q^{\mu}(b)$ by using~(\ref{eq:rel-constraint}) and~(\ref{eq:rel-sol-1}) evaluated at $\tau = b$. The result is
\begin{align}
&p_t =-\sigma\sqrt{\vec{p}^2c^2+m^2c^4} \ \ , \ \ \sigma = \pm1 \ , \label{eq:rel-momenta-boundary-1}\\
&\left(1+\frac{\vec{p}^2}{m^2c^2}\right)^{-1} = 1-\frac{(\vec{q}(b)-\vec{q}(a))^2}{c^2(t(b)-t(a))^2} \ .\label{eq:rel-momenta-boundary-2}
\end{align}
Equation~(\ref{eq:rel-momenta-boundary-1}) together with~(\ref{eq:rel-velo}) implies that $\mathrm{sgn}(\dot{t}) = -\mathrm{sgn}(p_t) = \sigma = \text{const}$, which leads to $|t(b)-t(a)| = \sigma(t(b)-t(a))$. Using~(\ref{eq:rel-momenta-boundary-1}) and~(\ref{eq:rel-momenta-boundary-2}), we can now insert the relational solution~(\ref{eq:rel-sol-1}) in the integrand of~(\ref{eq:rel-action}) to obtain the on-shell action
\begin{align*}
&W(ct(b),\vec{q}(b);ct(a),\vec{q}(a)) := I_{\text{on-shell}}\\
& = -\frac{\sigma m c^2}{\sqrt{1+\frac{\vec{p}^2}{m^2c^2}}}(t(b)-t(a))\\
&= -m c^2|t(b)-t(a)|\sqrt{1-\frac{(\vec{q}(b)-\vec{q}(a))^2}{c^2(t(b)-t(a))^2}} \ ,
\end{align*}
which can be rewritten as
\begin{equation}\label{eq:rel-on-shell-action}
\begin{aligned}
&W(ct(b),\vec{q}(b);ct(a),\vec{q}(a))\\ &= -mc\sqrt{c^2(t(b)-t(a))^2-(\vec{q}(b)-\vec{q}(a))^2} \ .
\end{aligned}
\end{equation}
This is the expected result from elementary relativity. One can readily verify that the on-shell action given in~(\ref{eq:rel-on-shell-action}) is a solution to the Hamilton-Jacobi (HJ) constraint,
\begin{equation}\label{eq:rel-HJ}
-\frac{1}{2c^2}\left(\frac{\partial W}{\partial t(b)}\right)^2+\frac{1}{2}\left(\frac{\partial W}{\partial\vec{q}(b)}\right)^2+\frac{m^2c^2}{2} = 0 \ ,
\end{equation}
similarly for the other end point. The unobservable label $\tau$ does not appear in these equations. Nevertheless, it is clear that the absence of ``time'' in~(\ref{eq:rel-HJ}) does not imply the absence of dynamics, but rather it signals that the dynamics is relational~\cite{Rovelli:2011,Rovelli:2014}. It is worthwhile to note that the same disappearance of label time occurs in the quantum theory of gauge-invariant states. This does not imply that there is no quantum dynamics or that the dynamics can only be understood in a particular (semiclassical) regime~(which is the view taken in the semiclassical approach to the problem of time). As in the classical theory, the dynamics has to be understood in a relational way.\footnote{In~\cite{Chataig:2019}, it was argued that the notion of ``gauge fixing'' (i.e., the fixation of the time coordinate) is fundamental to the understanding of the dynamics in a time-reparametrization invariant system. This means that the unobservable label $\tau$ can be chosen in order to describe the evolution of the dynamical fields in a relational way. This becomes clear when one considers the construction of invariant extensions through a gauge condition, as was done in Secs.~\ref{sec:observables-classical} and~\ref{sec:quantum-Dobs} of the present article.} We will see how this occurs when computing matrix elements of quantum Dirac observables. 

\subsubsection{Nonrelativistic limit}
Since we will be interested in the nonrelativistic limit of quantum Dirac observables, it is worthwhile to briefly discuss the classical setting. Using~(\ref{eq:rel-momenta-boundary-1}), we can expand~(\ref{eq:rel-DObs-1}) and~(\ref{eq:rel-DObs-2}) in a (formal) power series in $\frac{1}{c^2}$ to obtain
\begin{align}
\vec{q}(a) &= \vec{q}(\tau)-\frac{\sigma\vec{p}}{m\sqrt{1+\frac{\vec{p}^2}{m^2c^2}}}(t(\tau)-t(a))\notag\\
&=\vec{q}(\tau)-\frac{\sigma\vec{p}}{m}(t(\tau)-t(a))+\mathcal{O}\left(\frac{1}{c^2}\right) \ , \label{eq:nonrel-DObs-1}\\
t(a) &= t(\tau)-\frac{\sigma m\sqrt{1+\frac{\vec{p}^2}{m^2c^2}}}{p_1}(q^1(\tau)-q^1(a))\notag\\
&=t(\tau)-\frac{\sigma m}{p_1}(q^1(\tau)-q^1(a))+\mathcal{O}\left(\frac{1}{c^2}\right) \ , \label{eq:nonrel-DObs-2}\\
q^j(a) &= q^j(\tau)-\frac{p_j}{p_1}(q^1(\tau)-q^1(a)) \ . \label{eq:nonrel-DObs-3}
\end{align}
These are simply the Newtonian Dirac observables, which describe the relational evolution between $\vec{q}(\tau)$ and $t(\tau)$ in a gauge invariant manner. In particular, Eq.~(\ref{eq:nonrel-DObs-2}) is the time-of-arrival observable (see~\cite{Grot:1996,Hoehn:2018-2} and references therein), which corresponds to the value of $t(\tau)$ ``when'' $q^1(\tau) = q^1(a)$.  Similarly, the expansion of the on-shell action~(\ref{eq:rel-on-shell-action}) yields
\begin{align}
&W(ct(b),\vec{q}(b);ct(a),\vec{q}(a))\notag\\
& = -\sigma mc^2 (t(b)-t(a))\sqrt{1-\frac{(\vec{q}(b)-\vec{q}(a))^2}{c^2(t(b)-t(a))^2}}\notag\\
&= -\sigma mc^2 (t(b)-t(a))\left[1-\frac{(\vec{q}(b)-\vec{q}(a))^2}{2c^2(t(b)-t(a))^2}\right]+\mathcal{O}\left(\frac{1}{c^2}\right)\notag\\
&=: \varphi_{\sigma}(ct(b);ct(a)) + S_{\sigma}(t(b),\vec{q}(b);t(a),\vec{q}(a)) \ ,\label{eq:WKB-BO-classical-0}
\end{align}
where we defined
\begin{equation}\label{eq:WKB-BO-classical}
\varphi_{\sigma}(ct(b);ct(a)) := -\sigma mc^2 (t(b)-t(a)) \ 
\end{equation}
and
\begin{equation}\label{eq:WKB-BO-classical-2}
\begin{aligned}
&S_{\sigma}(t(b),\vec{q}(b);t(a),\vec{q}(a))\\
&:= \frac{\sigma m}{2} \frac{(\vec{q}(b)-\vec{q}(a))^2}{(t(b)-t(a))}+\mathcal{O}\left(\frac{1}{c^2}\right) \ .
\end{aligned}
\end{equation}
Up to order $c^0$, $S_{\sigma}$ solves the Newtonian HJ constraints
\begin{equation}\label{eq:nonrel-HJ}
\begin{aligned}
&+\sigma\frac{\partial S_{\sigma}}{\partial t(b)}+\frac{1}{2m}\left(\frac{\partial S_{\sigma}}{\partial\vec{q}(b)}\right)^2 = \mathcal{O}\left(\frac{1}{c^2}\right) \ ,\\
&-\sigma\frac{\partial S_{\sigma}}{\partial t(a)}+\frac{1}{2m}\left(\frac{\partial S_{\sigma}}{\partial\vec{q}(a)}\right)^2 = \mathcal{O}\left(\frac{1}{c^2}\right) \ ,
\end{aligned}
\end{equation}
i.e., $S_{\sigma}$ is the Newtonian on-shell action. When higher orders in $\frac{1}{c^2}$ are included, $S_{\sigma}$ solves corrected Newtonian constraints, where the corrections come from the expansion of the square root in~(\ref{eq:rel-momenta-boundary-1}). The same expansion procedure can be performed for any minisuperspace model of cosmology with a nonvanishing potential and, formally, for the field-theoretical case~\cite{Chataig:2019}. The corrected Newtonain constraint leads to a corrected Schr\"odinger equation in the quantum theory. In~\cite{Kiefer:1991}, the same formal procedure was applied to quantum geometrodynamics and lead to a corrected functional Schr\"odinger equation for both gravitational and matter fields. As was argued in~\cite{Chataig:2019}, the corrections do not violate unitarity of the evolution with respect to the relational time $t(\tau)$.

\subsection{The semiclassical approach to the problem of time: WKB time}
In any version of the quantum theory of a generally covariant system, the fundamental equation is the quantum constraint equation. For the model at hand, it can be obtained by promoting the classical constraint~(\ref{eq:rel-constraint}) to the wave equation
\begin{equation}\label{eq:WKB-KG}
\frac{\hbar^2}{2c^2}\frac{\partial^2\psi}{\partial t^2}-\frac{\hbar^2}{2}\frac{\partial^2\psi}{\partial\vec{q}^2}+\frac{m^2c^2}{2}\psi = 0 \ ,
\end{equation}
via the canonical quantization procedure, $p_t\to-\I\hbar\frac{\partial}{\partial t}, \vec{p}\to-\I\hbar\frac{\partial}{\partial\vec{q}}$. As in the classical case, the quantum constraint equation does not depend on the worldline label time $\tau$. This has led to some confusion regarding the dynamics of the wave function $\psi$.

One particular attitude toward the absence of $\tau$ (the problem of time) is that one can only define evolution in a particular (semiclassical) limit of the theory~\cite{Kuchar:1991,Isham:1992,Kiefer:1993-1,Kiefer:1993-2}. This is the so-called semiclassical approach to the problem of time. The adjective ``semiclassical'' does not refer to a formal expansion in powers of $\hbar$, but rather to a perturbative procedure in which one part of the system, called the ``heavy'' part, behaves classically while the other part, called the ``light'' part, can still be described quantum mechanically. As was argued in~\cite{Chataig:2019}, this procedure corresponds to a weak-coupling expansion, where the heavy part serves as a background with respect to which the evolution of the light part (the perturbations on the background) is described.

In this weak-coupling expansion, a time parameter ``emerges.'' It can therefore be referred to as the semiclassical emergent time, although it is also sometimes called the WKB time~\cite{Zeh:1987,Kiefer:1993-1,Kiefer:1993-2}.  The WKB time is simply the time parameter of the background (heavy part), which corresponds to a particular (not unique) class of gauge choices for the worldline label time~\cite{Chataig:2019}. Although this class of gauges is singled out by the weak-coupling expansion, there is no preferred choice of label time to describe the relational evolution of quantum Dirac observables at the exact level (cf.~Sec.~\ref{sec:general-quantum}), as we will discuss in what follows. In particular, we will show for the example of the relativistic particle that the WKB time emerges in the weak-coupling (nonrelativistic) limit of invariant transition amplitudes.

Let us briefly illustrate the usual derivation of WKB time for~(\ref{eq:WKB-KG}). The expansion parameter is $\frac{1}{c^2}$ and it appears in conjunction with the field $t$, which thus will be the heavy or ``background'' variable. The key step is to assume that the wave function can be factorized as
\begin{equation}\label{eq:BO-ansatz}
\psi(ct,\vec{q}) = N(c,m,\hbar)\e^{\frac{\I}{\hbar}\varphi(ct)}\chi(t,\vec{q}) \ ,
\end{equation}
where $N(c, m, \hbar)$ is a normalization factor and $\varphi(ct)$ is a solution to the background HJ equation
\begin{equation}\label{eq:BO-background-HJ}
-\frac{1}{2c^2}\left(\frac{\partial\varphi}{\partial t}\right)^2+\frac{m^2c^2}{2} = 0 \ .
\end{equation}
Moreover, we assume that $\chi(t,\vec{q})$ can be formally expanded in powers of $\frac{1}{c^2}$: $\chi(t,\vec{q}) = \chi^{(0)}(t,\vec{q})+\mathcal{O}\left(\frac{1}{c^2}\right)$. By inserting~(\ref{eq:BO-ansatz}) into~(\ref{eq:WKB-KG}) and using~(\ref{eq:BO-background-HJ}), we obtain an equation for $\chi(t,\vec{q})$,
\begin{equation}\label{eq:BO-1}
\frac{\I\hbar}{c^2}\frac{\partial\varphi}{\partial t}\frac{\partial\chi}{\partial t}-\frac{\hbar^2}{2}\frac{\partial^2\chi}{\partial\vec{q}^2}+\frac{\hbar^2}{2c^2}\frac{\partial^2\chi}{\partial t^2}+\frac{\I\hbar}{2c^2}\frac{\partial^2\varphi}{\partial t^2}\chi = 0 \ .
\end{equation}
If we now define the WKB time derivative as\footnote{The factor of $\frac{1}{m}$ is included in~(\ref{eq:WKB-time-derivative}) such that $\tau$ has units of time.}
\begin{equation}\label{eq:WKB-time-derivative}
\frac{\partial}{\partial\tau} := -\frac{1}{mc^2}\frac{\partial\varphi}{\partial t}\frac{\partial}{\partial t} \ ,
\end{equation}
we can rewrite~(\ref{eq:BO-1}) in the Schr\"odinger-like form
\begin{equation}\label{eq:BO-2}
-\I\hbar m\frac{\partial\chi}{\partial\tau}-\frac{\hbar^2}{2}\frac{\partial^2\chi}{\partial\vec{q}^2} = -\frac{\hbar^2}{2c^2}\frac{\partial^2\chi}{\partial t^2}-\frac{\I\hbar}{2c^2}\frac{\partial^2\varphi}{\partial t^2}\chi \ .
\end{equation}
If the right-hand side vanishes, then~(\ref{eq:BO-2}) is exactly the Schr\"odinger equation for the fields $\vec{q}$. In this way, one has ``recovered'' time (the variable $\tau$) in an appropriate limit. For this simple example, one sees that this procedure simply corresponds to the nonrelativistic limit. This was already noted in~\cite{Kiefer:1993-2,Kiefer:1991}, but no connection with an exact quantum theory for the constraint equation~(\ref{eq:WKB-KG}) (cf.~Sec.~\ref{sec:general-quantum}) was established. This connection will become clear in the next section, where we show how the nonrelativistic limit of the gauge-invariant transition amplitude leads precisely to the factorization~(\ref{eq:BO-ansatz}).

We already know a solution to the background Hamilton-Jacobi equation~(\ref{eq:BO-background-HJ}). It is given by lowest order on-shell action $\varphi_{\sigma}(ct(b); ct(a))$ given in~(\ref{eq:WKB-BO-classical}), if we take $t(b)\equiv t$. In this case, the WKB time derivative reads
\begin{equation}
\begin{aligned}
 \frac{\partial}{\partial\tau}= -\frac{1}{mc^2}\frac{\partial\varphi_{\sigma}}{\partial t(b)}\frac{\partial}{\partial t(b)} &= +\sigma\frac{\partial}{\partial t(b)} \equiv \sigma\frac{\partial}{\partial t} \ , 
\end{aligned}
\end{equation}
and~(\ref{eq:BO-2}) becomes
\begin{equation}\label{eq:BO-3}
+\sigma\frac{\hbar}{\I} \frac{\partial\chi}{\partial t}-\frac{\hbar^2}{2m}\frac{\partial^2\chi}{\partial\vec{q}^2} = \mathcal{O}\left(\frac{1}{c^2}\right) \ ,
\end{equation}
which is the quantum version of the first equation in~(\ref{eq:nonrel-HJ}). In fact, the factorization of the on-shell action in~(\ref{eq:WKB-BO-classical-0}) is the classical version of the factorization of the wave function in~(\ref{eq:BO-ansatz}). The lowest order action $\varphi$ can be understood as a background action, whereas the WKB time derivative describes the evolution of the ``perturbations'' given by $S_{\sigma}$ or $\chi$ with respect to this background. This is a straightforward consequence of the weak-coupling (nonrelativistic) limit, which singles out the label time gauge $\tau = t(\tau)$. Thus, as was argued in~\cite{Chataig:2019}, all the results of the semiclassical approach to the problem of time coincide with a choice of gauge in a weak-coupling regime. However, in the exact theory (as described in Sec.~\ref{sec:general-quantum}), more general choices of gauge can be adopted, and there is no need to consider a weak-coupling expansion.

A couple of final remarks about the semiclassical approach are in order. First, in this approach, one often associates an inner product only with the perturbations described by $\chi(t,\vec{q})$, since the absence of label time $\tau$ in~(\ref{eq:WKB-KG}) is taken to signify that one cannot define a (conserved, positive definite) inner product for the full generally covariant theory~\cite{Kiefer:1993-2}. This is not correct~\cite{Rovelli:book}, since it is, in fact, possible to define a positive-definite inner product for the full theory, with respect to which the evolution of Dirac observables is unitary (cf.~Sec.~\ref{sec:general-quantum}). This will be illustrated in the following sections. Evidently, it remains to be seen whether a quantum theory (of gravity) based on this induced inner product is realized in nature.

Second, one often attributes importance to the presence of a first-order derivative in~(\ref{eq:BO-2}) and~(\ref{eq:BO-3}) in order to recover time. This corresponds to the linear derivative term in~(\ref{eq:nonrel-HJ}), which is a linear momentum term in the constraint equation. This linear term appears due to the weak-coupling (in this case, nonrelativistic) limit, but it is not at all necessary that the constraint of a generally covariant theory be written in such a form (often called ``deparametrized''~\cite{Kiefer:book}). Indeed, a theory with a constraint that is quadratic in the momenta can be quantized as in Sec.~\ref{sec:general-quantum} and as in what follows. The relational evolution, which is encoded in the correlation between field configurations, and Dirac observables can be defined in the same way, regardless of whether the constraints are in the deparametrized form. Furthermore, the treatment of constraints which are quadratic in the momenta can be made equivalent to the usual gauge theories of internal symmetries, which have constraints that are linear in the momenta~\cite{HT-Vergara:1992}.

\subsection{Nonrelativistic limit of the invariant transition amplitude}
The objective of the quantum theory is to quantize the constraint~(\ref{eq:rel-constraint}), which yields~(\ref{eq:WKB-KG}) and to compute the dynamics of operators corresponding to the Dirac observables. To achieve this, we now use the general formalism presented in Sec.~\ref{sec:general-quantum}. We consider the auxiliary Hilbert space $L^2(\mathbb{R}^{d+1}, \D ct\D^d q)$ of square integrable functions defined in the particle's configuration space. The auxiliary inner product is\footnote{If the fundamental length, mass, and time units are $L, M$, and $T$, respectively, the basic quantities have the following dimensions: $[q^{\mu}] = L, [p_{\mu}] = \frac{ML}{T}, [\hbar] = \frac{ML^2}{T}, [\ket{q^{\mu}}] = L^{-\frac{1}{2}}, [\ket{p_{\mu}}] = \frac{T^{\frac{1}{2}}}{L^{\frac{1}{2}}M^{\frac{1}{2}}}$. If $\ket{\psi}$ is normalized in the kinematical inner product, $\braket{\psi|\psi} = 1$, then we have $ [\ket{\psi}] = 0$. Moreover, the Heaviside step function $\Theta(x)$ is dimensionless.}
\begin{equation}\notag
\braket{\psi^{(1)}|\psi^{(2)}} = \int_{-\infty}^{\infty}c\D t\int_{\mathbb{R}^d}\D^d q\ \bar{\psi}^{(1)}(ct,\vec{q})\psi^{(2)}(ct,\vec{q}) \ .
\end{equation}
The eigenstates of the constraint operator,
\begin{equation}
\op{C} = -\frac{1}{2c^2}\op{p}_t^2+\frac{1}{2}\op{\vec{p}}^2+\frac{m^2c^2}{2} \ ,
\end{equation}
can be written as
\begin{equation}\label{eq:rel-eigenstates}
\begin{aligned}
&\braket{ct,\vec{q}|\sigma, \vec{p},mc}\\ &= \frac{1}{(2\pi\hbar)^{\frac{d+1}{2}}}\exp\left(-\frac{\I}{\hbar}\sigma\sqrt{\vec{p}^2c^2+m^2c^4}t\right)\e^{\frac{\I}{\hbar}\vec{p}\cdot\vec{q}} \ ,
\end{aligned}
\end{equation}
where $\sigma = \pm1$ labels the positive and negative frequency sectors, respectively. The states given in~(\ref{eq:rel-eigenstates}) obey the normalization condition
\begin{equation}\label{eq:rel-kinematical-overlap}
\begin{aligned}
&\braket{\sigma',\vec{p}', m'c|\sigma, \vec{p},mc} = \sqrt{\vec{p}^2+m^2c^2}\\ &\times\delta_{\sigma',\sigma}\delta(\vec{p}'-\vec{p})\delta\left(\frac{m'^2c^2}{2}-\frac{m^2c^2}{2}\right) \ .
\end{aligned}
\end{equation}
The induced inner product for states on the same mass shell is (cf.~Sec.~\ref{sec:general-quantum})~\cite{HT-SUSY:1982,Marolf:1995-1}
\begin{equation}
\begin{aligned}
&(\sigma', \vec{p}', mc| \sigma, \vec{p}, mc)\\ &:= \sqrt{\vec{p}^2+m^2c^2}\ \delta_{\sigma',\sigma}\delta(\vec{p}'-\vec{p}) \ ,
\end{aligned}
\end{equation}
and the improper projector onto a given mass shell with a definite frequency can be defined as\footnote{Note that $\op{P}_{\sigma, m}$, as defined in~(\ref{eq:improper-projector}), has units of inverse momentum squared, $\left[\op{P}_{\sigma, m}\right] = \left(\frac{ML}{T}\right)^{-2}$. Thus, a state that is normalized in the induced inner product must have dimensions of momentum.}
\begin{equation}\label{eq:improper-projector}
\begin{aligned}
\op{P}_{\sigma, m}:= \int_{\mathbb{R}^d}\frac{\D^d p}{\sqrt{\vec{p}^2+m^2c^2}}\ket{\sigma,\vec{p},mc}\bra{\sigma,\vec{p},mc} \ .
\end{aligned}
\end{equation}
Using~(\ref{eq:rel-kinematical-overlap}), it is straightforward to verify that $\op{P}_{\sigma, m}$ satisfies
\begin{equation}\label{eq:rel-P-overlap}
\op{P}_{\sigma',m'}\op{P}_{\sigma,m} = \delta_{\sigma',\sigma}\delta\left(\frac{m'^2c^2}{2}-\frac{m^2c^2}{2}\right)\op{P}_{\sigma,m} \ .
\end{equation}
and that it has the matrix elements
\begin{align}
 &\left<\frac{p'_t}{c},\vec{p}'\left|\op{P}_{\sigma,m}\left|\frac{p_t}{c},\vec{p}\right.\right.\right> = \delta\left(\frac{p'_t}{c}-\frac{p_t}{c}\right)\delta(\vec{p}'-\vec{p})\notag\\
 &\times\delta\left(-\frac{p_t^2}{2c^2}+\frac{\vec{p}^2}{2}+\frac{m^2c^2}{2}\right)\Theta\left(-\frac{\sigma p_t}{c}\right) \ . \label{eq:matrix-P-momentum}
\end{align}
We can use this improper projector to extract the gauge-invariant part of the configuration eigenstates:
\begin{equation}
\ket{ct,\vec{q};\sigma,m} := \op{P}_{\sigma,m}\ket{ct,\vec{q}} \ .
\end{equation}
Let us now examine how the nonrelativistic expansion of the invariant transition amplitude leads to the factorization~(\ref{eq:BO-ansatz}) used in the semiclassical approach. As is well-known, the quantum analogue of the classical on-shell action~(\ref{eq:rel-on-shell-action}) is the transition amplitude between the gauge-invariant states,
\begin{equation}
\begin{aligned}
&(c t', \vec{q}';\sigma',m|ct,\vec{q};\sigma,m)\\
& = \delta_{\sigma', \sigma}\braket{ct',\vec{q}'|\op{P}_{\sigma,m}|ct,\vec{q}} \ .
\end{aligned}
\end{equation}
This amplitude is a solution to the constraint equation as can easily be verified. Let us now perform a formal expansion in powers of $\frac{1}{c^2}$ (weak coupling expansion; nonrelativistic limit). From~(\ref{eq:improper-projector}), we obtain
\begin{align}
&(ct', \vec{q}';\sigma',m| ct,\vec{q};\sigma,m)= \frac{\delta_{\sigma',\sigma}}{(2\pi\hbar)^{d+1}mc}\notag\\
&\times\int_{\mathbb{R}^d}\frac{\D^d p}{\sqrt{1+\frac{\vec{p}^2}{m^2c^2}}}\ \e^{-\frac{\I}{\hbar}\sigma mc^2\sqrt{1+\frac{\vec{p}^2}{m^2c^2}}(t'-t)}\e^{\frac{\I}{\hbar}\vec{p}\cdot(\vec{q}'-\vec{q})}\notag\\
&= \frac{\delta_{\sigma',\sigma}}{2\pi\hbar mc}\e^{-\frac{\I}{\hbar}\sigma mc^2(t'-t)}\int_{\mathbb{R}^d}\frac{\D^d p}{(2\pi\hbar)^d}\ \e^{-\frac{\I}{\hbar}\sigma\frac{\vec{p}^2}{2m}(t'-t)}\notag\\
&\times\e^{\frac{\I}{\hbar}\vec{p}\cdot(\vec{q}'-\vec{q})}+ \mathcal{O}\left(\frac{1}{c^3}\right)\notag\\
&= \frac{\delta_{\sigma',\sigma}}{2\pi\hbar mc} \e^{\frac{\I}{\hbar}\varphi_{\sigma}(ct';ct)}K_{\sigma}(t',\vec{q}';t,\vec{q}) + \mathcal{O}\left(\frac{1}{c^3}\right)\ ,\label{eq:transition-BO}
\end{align}
where $\varphi_{\sigma}(ct';ct)$ is the lowest-order on-shell action defined in~(\ref{eq:WKB-BO-classical}) and
\begin{equation}\label{eq:nonrel-propagator-1}
\begin{aligned}
&K_{\sigma}(t',\vec{q}';t,\vec{q})\\ &= \left(\frac{m}{2\pi\I\hbar\sigma(t'-t)}\right)^{\frac{d}{2}} \exp\left(-\frac{m(\vec{q}'-\vec{q})^2}{2\I\hbar\sigma(t'-t)}\right)
\end{aligned}
\end{equation}
is the nonrelativistic propagator, which is a solution to the Schr\"odinger constraint. The overall factor of $\frac{1}{\hbar mc}$ in~(\ref{eq:transition-BO}) appears for dimensional reasons. Equation~(\ref{eq:transition-BO}) shows that all the results of the semiclassical approach to the problem of time can be recovered from the exact quantum theory (cf.~Sec.~\ref{sec:general-quantum}) based on the induced inner product in the weak-coupling (nonrelativistic) limit. In Secs.~\ref{sec:rel-qDobs-1} and~\ref{sec:rel-qDobs-2}, we will illustrate how the relational evolution of quantum Dirac observables can be understood without resorting to this weak-coupling limit.

\subsection{\label{sec:rel-qDobs-1}Quantum observables I}
We now follow the general formalism presented in Sec.~\ref{sec:quantum-Dobs} to construct quantum Dirac observables for the relativistic particle. Let us consider the physical states $\ket{\frac{p_t}{c},\vec{p};\sigma,m}:=\op{P}_{\sigma,m}\ket{\frac{p_t}{c},\vec{p}}$. According to~(\ref{eq:physical-matrix-elements-DObs}), the matrix elements of a Dirac observable $\op{\mathcal{O}}_{\omega}$ are defined as
\begin{equation}
\begin{aligned}
&\left(\frac{p_t'}{c},\vec{p}';\sigma',m\left|\op{\mathcal{O}}_{\omega}\right|\frac{p_t}{c},\vec{p};\sigma,m\right)\\
&:= 2\pi\hbar\left<\frac{p_t'}{c},\vec{p}'\left|\op{P}_{\sigma',m}\op{\omega}\op{P}_{\sigma,m}\left|\frac{p_t}{c},\vec{p}\right.\right.\right> \ .
\end{aligned}
\end{equation}
In this way, we can compute the physical matrix elements of Dirac observables by inserting the operators
\begin{equation}\label{eq:observable-practical}
\op{\mathcal{O}}^m_{\omega}:=2\pi\hbar\sum_{\sigma', \sigma}\op{P}_{\sigma',m}\op{\omega}\op{P}_{\sigma,m}
\end{equation}
into the auxiliary inner product of two states. The Faddeev-Popov resolution of the identity given in~(\ref{eq:FP-quantum-0}) can thus be written as
\begin{equation}\label{eq:FP-practical}
\begin{aligned}
&2\pi\hbar\left<\frac{p_t'}{c},\vec{p}'\left|\op{P}_{\sigma',m}\op{\omega}[1|\chi = s]\op{P}_{\sigma,m}\left|\frac{p_t}{c},\vec{p}\right.\right.\right>\\
& = \delta_{\sigma',\sigma}\left<\frac{p_t'}{c},\vec{p}'\left|\op{P}_{\sigma,m}\left|\frac{p_t}{c},\vec{p}\right.\right.\right> \ .
\end{aligned}
\end{equation}
\subsubsection{Matrix elements}
We first consider the gauge condition $c\op{t}$, which is self-adjoint with respect to the auxiliary inner product. To construct $\op{\omega}[1|ct= cs]$, we define
\begin{equation}
\op{\Delta}_{ct} :=-\frac{\I}{\hbar}[c\op{t},\op{C}] = -\frac{1}{c}\op{p}_t \ ,
\end{equation}
which happens to be already invariant, i.e., $[\op{p}_t,\op{C}] = 0$. Thus, we have $\op{\Delta}_{ct}^{\mathcal{O}} \equiv \op{\Delta}_{ct}$~[cf.~(\ref{eq:FP-quantum-inverse-measure})]. We can then define~[cf.~(\ref{eq:general-relational-scalar-quantum}) and~(\ref{eq:observable-practical})]
\begin{equation}\label{eq:ket-omega-1}
\ket{\sigma, \vec{q};s} := \left|\frac{\op{p}_t}{c}\right|^{\frac{1}{2}}\Theta\left(-\frac{\sigma\op{p}_t}{c}\right)\ket{ct = cs,\vec{q}} \ ,
\end{equation}
and
\begin{equation}\label{eq:omega-f-q}
\begin{aligned}
&\op{\omega}\left[f(\vec{q})|ct = cs\right]\\&:=\sum_{\sigma = \pm}\int_{\mathbb{R}^d}\D^dq\ f(\vec{q})\ket{\sigma, \vec{q};s}\bra{\sigma, \vec{q};s} \ .
\end{aligned}
\end{equation}
It is straightforward to verify that~(\ref{eq:FP-practical}) is satisfied.
Using~(\ref{eq:ket-omega-1}) and~(\ref{eq:omega-f-q}), we find
\begin{widetext}
\begin{align*}
&2\pi\hbar\left<\frac{p_t'}{c},\vec{p}'\left|\op{P}_{\sigma',m}\op{\omega}\left[1|ct = cs\right]\op{P}_{\sigma,m}\left|\frac{p_t}{c},\vec{p}\right.\right.\right>=\delta_{\sigma',\sigma}\int_{\mathbb{R}^d}\frac{\D^dq}{(2\pi\hbar)^{d}}\ \Theta\left(-\frac{\sigma p_t'}{c}\right)\Theta\left(-\frac{\sigma p_t}{c}\right)\left|\frac{p_t' p_t}{c^2}\right|^{\frac{1}{2}}\\
&\times\e^{\frac{\I}{\hbar}s(p_t-p_t')}\e^{\frac{\I}{\hbar}\vec{q}\cdot(\vec{p}-\vec{p}')}\delta\left(-\frac{p_t'^2}{2c^2}+\frac{\vec{p}'^2}{2}+\frac{m^2c^2}{2}\right)\delta\left(-\frac{p_t^2}{2c^2}+\frac{\vec{p}^2}{2}+\frac{m^2c^2}{2}\right)\ .
\end{align*}
After integrating over $\vec{q}$, this yields
\begin{align*}
&2\pi\hbar\left<\frac{p_t'}{c},\vec{p}'\left|\op{P}_{\sigma',m}\op{\omega}\left[1|ct = cs\right]\op{P}_{\sigma,m}\left|\frac{p_t}{c},\vec{p}\right.\right.\right>\\
&=\delta_{\sigma',\sigma} \Theta\left(-\frac{\sigma p_t'}{c}\right)\Theta\left(-\frac{\sigma p_t}{c}\right)\left|\frac{p_t' p_t}{c^2}\right|^{\frac{1}{2}}\e^{\frac{\I}{\hbar}s(p_t-p_t')}\delta(\vec{p}-\vec{p}')\delta\left(-\frac{p_t'^2}{2c^2}+\frac{p_t^2}{2c^2}\right)\delta\left(-\frac{p_t^2}{2c^2}+\frac{\vec{p}^2}{2}+\frac{m^2c^2}{2}\right)\\
&=\delta_{\sigma',\sigma} \delta\left(-\frac{p_t^2}{2c^2}+\frac{\vec{p}^2}{2}+\frac{m^2c^2}{2}\right)\delta\left(\frac{p_t'}{c}-\frac{p_t}{c}\right)\delta(\vec{p}'-\vec{p})\Theta\left(-\frac{\sigma p_t}{c}\right)=\delta_{\sigma',\sigma}\left<\frac{p_t'}{c},\vec{p}'\left|\op{P}_{\sigma,m}\left|\frac{p_t}{c},\vec{p}\right.\right.\right>\ ,
\end{align*}
where we used~(\ref{eq:matrix-P-momentum}). A similar calculation yields~[cf.~(\ref{eq:omega-f-q})]
\begin{equation}\label{eq:rel-DObs-1-quantum-matrix}
\begin{aligned}
&2\pi\hbar\left<\frac{p_t'}{c},\vec{p}'\left|\op{P}_{\sigma',m}\op{\omega}\left[\vec{q}|ct = cs\right]\op{P}_{\sigma,m}\left|\frac{p_t}{c},\vec{p}\right.\right.\right>=\delta_{\sigma',\sigma}\Theta\left(-\frac{\sigma p_t'}{c}\right)\Theta\left(-\frac{\sigma p_t}{c}\right)\left|\frac{p_t' p_t}{c^2}\right|^{\frac{1}{2}}\\
&\times\e^{\frac{\I}{\hbar}s(p_t-p_t')}\left[\frac{\hbar}{\I}\frac{\partial}{\partial\vec{p}}\delta(\vec{p}-\vec{p}')\right]\delta\left(-\frac{p_t'^2}{2c^2}+\frac{\vec{p}'^2}{2}+\frac{m^2c^2}{2}\right)\delta\left(-\frac{p_t^2}{2c^2}+\frac{\vec{p}^2}{2}+\frac{m^2c^2}{2}\right) \ .
\end{aligned}
\end{equation}
Equation~(\ref{eq:rel-DObs-1-quantum-matrix}) gives the matrix elements of the invariant extension of $\op{\vec{q}}$ with respect to the gauge condition $c\op{t}$.
\end{widetext}

\subsubsection{Relation to the classical expression}
Can we relate the matrix elements given in~(\ref{eq:rel-DObs-1-quantum-matrix}) to the classical expression given in~(\ref{eq:rel-DObs-1})? The answer is yes. To see this, let us choose two test functions $\psi^{(1,2)}\left(\frac{p_t}{c},\vec{p}\right)$ with compact support in momentum space. Furthermore, we require that $\psi^{(1,2)}(0,\vec{p}) = 0$ and we define 
\begin{equation}\notag
\psi^{(1,2)}_{\sigma}(\vec{p}):=\psi^{(1,2)}\left(-\sigma\sqrt{\vec{p}^2+m^2c^2},\vec{p}\right) \ 
\end{equation}
for brevity. From~(\ref{eq:rel-DObs-1-quantum-matrix}), we obtain
\begin{equation}\label{eq:compact-support-DObs-1-1}
\begin{aligned}
&2\pi\hbar\sum_{\sigma', \sigma}\left<\psi^{(1)}\left|\op{P}_{\sigma',m}\op{\omega}\left[\vec{q}|ct = cs\right]\op{P}_{\sigma,m}\left|\psi^{(2)}\right.\right.\right>\\
&=\frac{\hbar}{\I}\sum_{\sigma = \pm}\int\frac{\D^d p'\D^d p\ \bar{\psi}^{(1)}_{\sigma}(\vec{p}')\psi^{(2)}_{\sigma}(\vec{p})}{\left(\vec{p}'^2+m^2c^2\right)^{\frac{1}{4}}\left(\vec{p}^2+m^2c^2\right)^{\frac{1}{4}}}\\
&\times\e^{-\frac{\I}{\hbar}sc\sigma\left(\sqrt{\vec{p}^2+m^2c^2}-\sqrt{\vec{p}'^2+m^2c^2}\right)}\frac{\partial}{\partial\vec{p}}\delta(\vec{p}-\vec{p}')\\
&=\sum_{\sigma = \pm}\int_{\mathbb{R}^d}\frac{\D^d p}{\sqrt{\vec{p}^2+m^2c^2}}\ \bar{\psi}^{(1)}_{\sigma}(\vec{p})\left[\I\hbar\frac{\partial}{\partial\vec{p}}\right.\\
&\left.+\frac{c\vec{p}}{\sigma\sqrt{\vec{p}^2+m^2c^2}}s-\frac{\I\hbar\vec{p}}{2\left(\vec{p}^2+m^2c^2\right)}\right]\psi^{(2)}_{\sigma}(\vec{p}) \ .
\end{aligned}
\end{equation}
We now observe that
\begin{align*}
&\frac{\partial}{\partial\vec{p}}\psi^{(1,2)}_{\sigma}(\vec{p}) = \left.\frac{\partial}{\partial\vec{p}}\psi^{(1,2)}\left(\frac{p_t}{c},\vec{p}\right)\right|_{\frac{p_t}{c} = -\sigma\sqrt{\vec{p}^2+m^2c^2}}\\
&+\left.\frac{c^2\vec{p}}{p_t}\frac{\partial}{\partial p_t}\psi^{(1,2)}\left(\frac{p_t}{c},\vec{p}\right)\right|_{\frac{p_t}{c} = -\sigma\sqrt{\vec{p}^2+m^2c^2}} \ ,
\end{align*}
such that~(\ref{eq:compact-support-DObs-1-1}) can be written as
\begin{equation}\notag
\begin{aligned}
&2\pi\hbar\sum_{\sigma', \sigma}\left<\psi^{(1)}\left|\op{P}_{\sigma',m}\op{\omega}\left[\vec{q}|ct = cs\right]\op{P}_{\sigma,m}\left|\psi^{(2)}\right.\right.\right>\\
&=\int\frac{\D p_t}{c}\D^d p\ \bar{\psi}^{(1)}\left(\frac{p_t}{c},\vec{p}\right)\delta\left(-\frac{p_t^2}{2c^2}+\frac{\vec{p}^2}{2}+\frac{m^2c^2}{2}\right)\\
&\times\left[\I\hbar\frac{\partial}{\partial\vec{p}}+\I\hbar\frac{c^2\vec{p}}{p_t}\frac{\partial}{\partial p_t}-\frac{c^2\vec{p}}{p_t}s-\I\hbar\frac{c^2\vec{p}}{2p_t^2}\right]\psi^{(2)}\left(\frac{p_t}{c},\vec{p}\right) \ .
\end{aligned}
\end{equation}
Thus, the matrix elements of the invariant extension of $\op{\vec{q}}$ with respect to the gauge condition $c\op{t}$ coincide with the insertion of the operator
\begin{equation}\label{eq:insertion-Dobs-q}
\I\hbar\frac{\partial}{\partial\vec{p}}+\I\hbar\frac{c^2\vec{p}}{p_t}\frac{\partial}{\partial p_t}-\frac{c^2\vec{p}}{p_t}s-\I\hbar\frac{c^2\vec{p}}{2p_t^2}
\end{equation}
into the momentum-space induced inner product of the two test functions $\psi^{(1,2)}\left(\frac{p_t}{c},\vec{p}\right)$. It is straightforward to check that the operator given in~(\ref{eq:insertion-Dobs-q}) is symmetric in the auxiliary inner product and commutes with the constraint operator. Thus, it is symmetric in the induced inner product. It corresponds to a symmetric quantization of the classical Dirac observable given in~(\ref{eq:rel-DObs-1}), provided one makes the identifications $\vec{q}(\tau) \to \I\hbar\frac{\partial}{\partial\vec{p}}, t(\tau)\to \I\hbar\frac{\partial}{\partial p_t}, t(a) \to s$.

\subsubsection{Dynamics}
From~(\ref{eq:observable-practical}), we see that the physical eigenstates of the invariant extension $\op{\mathcal{O}}_m\left[\vec{q}|ct = cs\right] := 2\pi\hbar\sum_{\sigma',\sigma}\op{P}_{\sigma',m}\op{\omega}\left[\vec{q}|ct =c s\right]\op{P}_{\sigma,m}$ with eigenvalues $\vec{q}$ are given by the gauge-invariant component of the states given in~(\ref{eq:ket-omega-1}), i.e., $\ket{\sigma, \vec{q}; s,m}:=\sqrt{2\pi\hbar}\sum_{\sigma'}\op{P}_{\sigma',m}\ket{\sigma, \vec{q}; s}$, which can be written as
\begin{equation}\label{eq:eigenstates-insertion-Dobs-q}
\begin{aligned}
&\left<\left.\frac{p_t}{c},\vec{p}\right|\sigma, \vec{q};s,m\right>= \frac{1}{(2\pi\hbar)^{\frac{d}{2}}}\Theta\left(-\frac{\sigma p_t}{c}\right)\left|\frac{p_t}{c}\right|^{\frac{1}{2}}\\
&\times\e^{-\frac{\I}{\hbar}s p_t}\e^{-\frac{\I}{\hbar}\vec{q}\cdot\vec{p}}\delta\left(-\frac{p_t^2}{2c^2}+\frac{\vec{p}^2}{2}+\frac{m^2c^2}{2}\right) \ .
\end{aligned}
\end{equation}
A similar expression to the one given in~(\ref{eq:eigenstates-insertion-Dobs-q}) was also found in~\cite{Gambini:2001-1} from a direct computation of the eigenvalue problem for the operator given in~(\ref{eq:insertion-Dobs-q}). While this is certainly acceptable, for more general models it may be difficult to solve the classical equations of motion and, thus, to find classical expressions for the Dirac observables which are subsequently quantized. The method we have presented in Sec.~\ref{sec:quantum-Dobs}, which we exemplify here, is more general and can be applied to any generally covariant quantum-mechanical model, provided the eigenstates of the constraint operator are known. By focusing on the construction of the observables via their spectral decomposition [cf.~(\ref{eq:general-relational-scalar-quantum})], the method avoids the need to explicitly compute the classical Dirac observables beforehand.\footnote{Evidently, the computation of the inverse Faddeev-Popov invariant measure in~(\ref{eq:FP-quantum-inverse-measure}) requires the integration over $\tau$, which in principle implies that one would need to know the solutions to the (Heisenberg) equations of motion. However, it is sufficient to determine only the matrix elements of $\op{\Delta}^{\mathcal{O}}_{\chi}$ between physical states. For this, it is only necessary to know the solutions to $\op{C}\ket{\psi} = 0$, and it is not necessary to integrate over $\tau$.}

It is straightforward to verify that the states given in~(\ref{eq:eigenstates-insertion-Dobs-q}) evolve unitarily in $s$ and that their evolution is generated by the Dirac observable $\op{p}_t$:
\begin{equation}
\I\hbar\frac{\partial}{\partial s}\ket{\sigma, \vec{q}; s,m} = \op{p}_t\ket{\sigma, \vec{q};s,m} \ .
\end{equation}
Thus, the Dirac observable~(cf.~(\ref{eq:observable-practical}))
\begin{equation}\label{eq:observable-practical-1}
\begin{aligned}
&\op{\mathcal{O}}_m[\vec{q}|ct = cs]\\
&=\sum_{\sigma = \pm}\int_{\mathbb{R}^d}\D^dq\ \vec{q}\ket{\sigma,\vec{q};s,m}\bra{\sigma,\vec{q};s,m}
\end{aligned}
\end{equation}
obeys the Heisenberg equation of motion
\begin{equation}
\I\hbar\frac{\partial}{\partial s}\op{\mathcal{O}}_m[\vec{q}|ct = cs] = \left[\op{p}_t, \op{\mathcal{O}}_m[\vec{q}|ct = cs]\right]\ ,
\end{equation}
which is the quantum analogue of the classical equation~(\ref{eq:PB-classical-1}).

\subsubsection{Nonrelativistic limit}
Let us now compute the nonrelativistic limit of the matrix elements of~(\ref{eq:observable-practical-1}). Using~(\ref{eq:eigenstates-insertion-Dobs-q}), we find
\begin{align*}
&\left<\left.ct,\vec{q}\right|\sigma, \tilde{\vec{q}};s,m\right>=\frac{\e^{-\frac{\I}{\hbar}\sigma mc^2(t-s)}}{\sqrt{2\pi\hbar mc}}\\
&\times\int_{\mathbb{R}^d}\frac{\D^d p}{(2\pi\hbar)^{d}}\e^{-\frac{\I}{\hbar}\sigma \frac{\vec{p}^2}{2m}(t-s)}\e^{\frac{\I}{\hbar}\vec{p}\cdot(\vec{q}-\tilde{\vec{q}})}+\mathcal{O}\left(\frac{1}{c^{\frac{5}{2}}}\right)\\
&= \frac{\e^{-\frac{\I}{\hbar}\sigma mc^2(t-s)}}{\sqrt{2\pi\hbar mc}}K_{\sigma}(t,\vec{q};s,\tilde{\vec{q}})+\mathcal{O}\left(\frac{1}{c^{\frac{5}{2}}}\right) \ ,
\end{align*}
where $K_{\sigma}(t,\vec{q};s,\tilde{\vec{q}})$ is the nonrelativistic propagator given in~(\ref{eq:nonrel-propagator-1}). Thus, the nonrelativistic limit of the matrix elements of~(\ref{eq:observable-practical-1}) reads
\begin{align*}
&\left<ct',\vec{q}'\left|\op{\mathcal{O}}_m[\vec{q}|ct = cs]\right|ct,\vec{q}\right>\\
&=\frac{1}{2\pi\hbar mc}\sum_{\sigma=\pm}\e^{-\frac{\I}{\hbar}\sigma mc^2(t'-t)}\\
&\times\int_{\mathbb{R}^d}\D^d \tilde{q}\ \tilde{\vec{q}}\ K_{\sigma}(t',\vec{q}';s,\tilde{\vec{q}})K_{\sigma}(s,\tilde{\vec{q}};t,\vec{q})+\mathcal{O}\left(\frac{1}{c^{3}}\right) \ ,
\end{align*}
i.e., we recover the Newtonian matrix elements up to a WKB phase~(cf.~(\ref{eq:transition-BO})). Furthermore, the Newtonian matrix elements can be related to the classical Newtonian Dirac observable given in~(\ref{eq:nonrel-DObs-1}) as follows. As is well-known~\cite{Rovelli:book}, the nonrelativistic propagator can be understood as the Newtonian invariant transition amplitude,
\begin{equation}\label{eq:nonrel-propagator-2}
K_{\sigma}(t',\vec{q}';t,\vec{q}) = 2\pi\hbar\braket{t',\vec{q}'|\op{P}_{\sigma,m}^{\text{nonrel}}|t,\vec{q}} \ ,
\end{equation}
where
\begin{equation}\label{eq:nonrel-propagator-projector}
\op{P}_{\sigma,m}^{\text{nonrel}} = \int_{-\infty}^{\infty}\frac{\D\tau}{2\pi\hbar}\ \exp\left[\frac{\I}{\hbar}\tau\left(\sigma\op{p}_t+\frac{1}{2m}\op{\vec{p}}^2\right)\right]
\end{equation}
is the improper projector onto solutions of the Schr\"odinger constraint [cf.~(\ref{eq:BO-3})]. We may therefore write
\begin{align*}
&\frac{1}{2\pi\hbar}\int_{\mathbb{R}^d}\D^d \tilde{q}\ \tilde{\vec{q}} K_{\sigma}(t',\vec{q}';s,\tilde{\vec{q}})K_{\sigma}(s,\tilde{\vec{q}};t,\vec{q})\\
&= \left<t',\vec{q}'\left|\op{\mathcal{O}}_{\sigma,m}^{\text{nonrel}}[\vec{q}|t = s]\right|t,\vec{q}\right> \ ,
\end{align*}
where~[cf.~(\ref{eq:observable-practical-1})]
\begin{equation}\label{eq:observable-practical-nonrel-1}
\begin{aligned}
&\frac{1}{2\pi\hbar}\op{\mathcal{O}}_{\sigma,m}^{\text{nonrel}}[\vec{q}|t = s]\\
&:=\int\D^d\tilde{q}\ \tilde{\vec{q}}\ \op{P}_{\sigma,m}^{\text{nonrel}}\ket{t = s,\tilde{\vec{q}}}\bra{t = s,\tilde{\vec{q}}}\op{P}_{\sigma,m}^{\text{nonrel}} \ .
\end{aligned}
\end{equation}
In analogy to the derivation of~(\ref{eq:insertion-Dobs-q}), we may now compute the matrix element of $\op{\mathcal{O}}_{\sigma,m}^{\text{nonrel}}[\vec{q}|t = s]$ between two test functions $\psi^{(1,2)}(p_t,\vec{p})$ of compact support. We obtain~[cf.~(\ref{eq:insertion-Dobs-q})]
\begin{equation}\label{eq:non-rel-insertion}
\begin{aligned}
&\left<\psi^{(1)}\left|\op{\mathcal{O}}_{\sigma,m}^{\text{nonrel}}[\vec{q}|t = s]\right|\psi^{(2)}\right>\\
&= \int\D p_t\D^dp\ \bar{\psi}^{(1)}(p_t,\vec{p})\delta\left(\sigma p_t+\frac{\vec{p}^2}{2m}\right)\\
&\times\left\{\I\hbar\frac{\partial}{\partial\vec{p}}-\I\hbar\frac{\sigma\vec{p}}{m}\frac{\partial}{\partial p_t}+\frac{\sigma\vec{p}s}{m}\right\}\psi^{(2)}(p_t,\vec{p}) \ ,
\end{aligned}
\end{equation}
which is a symmetric quantization of the classical Newtonian observable given in~(\ref{eq:nonrel-DObs-1}), provided we identify $\vec{q}(\tau)\to\I\hbar\frac{\partial}{\partial\vec{p}}, t(\tau)\to\I\hbar\frac{\partial}{\partial p_t}, t(a)\to s$. Thus, our definition of observables given in~(\ref{eq:observable-practical}), which was motivated from the discussion in Sec.~\ref{sec:quantum-Dobs}, reproduces the correct results, both in the relativistic case and in the nonrelativistic limit.

\subsection{\label{sec:rel-qDobs-2}Quantum observables II}
\subsubsection{Matrix elements}
We now repeat the analysis of the last section for the gauge condition $\op{q}^1$, which is self-adjoint with respect to the auxiliary inner product. As before, we define
\begin{align}
&\op{\Delta}_{q^1} :=-\frac{\I}{\hbar}[\op{q}^1,\op{C}] = \op{p}_1 \equiv \op{\Delta}^{\mathcal{O}}_{q^1} \ ,\\
&\ket{\sigma, t, q^j;s} := \left|\op{p}_1\right|^{\frac{1}{2}}\Theta\left(\sigma\op{p}_1\right)\ket{ct, q^1 = cs, q^j} \ ,\label{eq:ket-omega-2}
\end{align}
and
\begin{align*}
&\op{\omega}\left[1|q^1 = cs\right]\\
&:=\sum_{\sigma = \pm}\int_{-\infty}^{\infty}\D ct\int_{\mathbb{R}^{d-1}}\D^{d-1}q\ \ket{\sigma, t, q^j;s}\bra{\sigma, t, q^j;s} \ .
\end{align*}
These definitions imply that $2\pi\hbar\op{P}_{\sigma',m}\op{\omega}\left[1|q^1 = cs\right]\op{P}_{\sigma,m}$ resolves the identity in the physical Hilbert space~[cf.~(\ref{eq:FP-practical})]:
\begin{equation}\notag
\begin{aligned}
&2\pi\hbar\left<\frac{p_t'}{c},\vec{p}'\left|\op{P}_{\sigma',m}\op{\omega}\left[1|q^1 = cs\right]\op{P}_{\sigma,m}\left|\frac{p_t}{c},\vec{p}\right.\right.\right>\\
&=\sum_{\sigma'' = \pm}\delta\left(-\frac{p_t^2}{2c^2}+\frac{\vec{p}^2}{2}+\frac{m^2c^2}{2}\right)\Theta\left(-\frac{\sigma p_t}{c}\right)\Theta\left(-\frac{\sigma' p_t}{c}\right)\\
&\times\Theta(\sigma''p_1')\Theta(\sigma''p_1)\e^{\frac{\I}{\hbar}cs(p_1-p_1')}|p_1'p_1|^{\frac{1}{2}}\\
&\times\delta\left(\frac{p_1'^2}{2}-\frac{p_1^2}{2}\right)\delta(p_t-p_t')\prod_{j=2}^d\delta(p_j-p_j')\\
&=\left(\sum_{\sigma'' = \pm}\Theta(\sigma''p_1)\right)\delta_{\sigma',\sigma}\delta(p_t-p_t')\delta(\vec{p}-\vec{p}')\\
&\times\delta\left(-\frac{p_t^2}{2c^2}+\frac{\vec{p}^2}{2}+\frac{m^2c^2}{2}\right)\Theta\left(-\frac{\sigma p_t}{c}\right)\\
&=\delta_{\sigma',\sigma}\left<\frac{p_t'}{c},\vec{p}'\left|\op{P}_{\sigma,m}\left|\frac{p_t}{c},\vec{p}\right.\right.\right>\ .
\end{aligned}
\end{equation}
We wish to compute the matrix elements of the invariant extension of $\op{t}$ with respect to the gauge condition $\op{q}^1$. We begin by computing
\begin{widetext}
\begin{equation}
\begin{aligned}
&2\pi\hbar\left<\frac{p_t'}{c},\vec{p}'\left|\op{P}_{\sigma',m}\op{\omega}\left[t|q^1 = cs\right]\op{P}_{\sigma,m}\left|\frac{p_t}{c},\vec{p}\right.\right.\right>=\sum_{\sigma'' = \pm}\Theta\left(\sigma'' p_1'\right)\Theta\left(\sigma'' p_1\right)\left|p_1'p_1\right|^{\frac{1}{2}}\\
&\times\e^{\frac{\I}{\hbar}cs(p_1-p_1')}\left[\frac{\hbar}{\I}\frac{\partial}{\partial p_t}\delta\left(\frac{p_t}{c}-\frac{p_t'}{c}\right)\right]\left[\prod_{j = 2}^d\delta(p_j-p_j')\right]\Theta\left(-\frac{\sigma' p_t'}{c}\right)\Theta\left(-\frac{\sigma p_t}{c}\right)\\
&\times\delta\left(-\frac{p_t'^2}{2c^2}+\frac{\vec{p}'^2}{2}+\frac{m^2c^2}{2}\right)\delta\left(-\frac{p_t^2}{2c^2}+\frac{\vec{p}^2}{2}+\frac{m^2c^2}{2}\right) \ .
\end{aligned}
\end{equation}
\end{widetext}
As before, we can relate this to the classical expression given in the first line of~(\ref{eq:rel-DObs-2}) by evaluating the matrix element between two test functions $\psi^{(1,2)}\left(\frac{p_t}{c},\vec{p}\right)$ of compact support, which satisfy $\psi^{(1,2)}\left(0, \vec{p}\right) = \psi^{(1,2)}\left(\frac{p_t}{c}, p_1 = 0, p_j\right) = 0$. We find
\begin{align*}
&2\pi\hbar\sum_{\sigma', \sigma}\left<\psi^{(1)}\left|\op{P}_{\sigma', m}\op{\omega}\left[t|q^1 = cs\right]\op{P}_{\sigma,m}\left|\psi^{(2)}\right.\right.\right>\\
&=\int\frac{\D p_t }{c}\D^{d} p\ \bar{\psi}^{(1)}\left(\frac{p_t}{c},\vec{p}\right)\delta\left(-\frac{p_t^2}{2c^2}+\frac{\vec{p}^2}{2}+\frac{m^2c^2}{2}\right)\\
&\times\left\{\I\hbar\frac{\partial}{\partial p_t}+\I\hbar\frac{p_t}{c^2p_1}\frac{\partial}{\partial p_1}-\frac{p_t s}{c p_1}-\I\hbar\frac{p_t}{2c^2p_1^2}\right\}\psi^{(2)}\left(\frac{p_t}{c},\vec{p}\right) \ ,
\end{align*}
i.e., the matrix elements of the invariant extension of $\op{t}$ are given by the insertion of
\begin{equation}\label{eq:insertion-Dobs-t}
\I\hbar\frac{\partial}{\partial p_t}+\I\hbar\frac{p_t}{c^2p_1}\frac{\partial}{\partial p_1}-\frac{p_t s}{c p_1}-\I\hbar\frac{p_t}{2c^2p_1^2}
\end{equation}
into the momentum-space induced inner product of the two test functions. The operator given in~(\ref{eq:insertion-Dobs-t}) is symmetric and commutes with the constraint operator. It corresponds to a symmetric quantization of the classical Dirac observable given in the first line of~(\ref{eq:rel-DObs-2}) if one makes the identifications $t(\tau)\to\I\hbar\frac{\partial}{\partial p_t}, q^1(\tau)\to\I\hbar\frac{\partial}{\partial p_1}, q^1(a)\to cs$.

\subsubsection{Dynamics}
The physical eigenstates of the invariant extension $\op{\mathcal{O}}_m[t|q^1= cs] := 2\pi\hbar\sum_{\sigma', \sigma}\op{P}_{\sigma',m}\op{\omega}[t|q^1 = cs]\op{P}_{\sigma,m}$ with eigenvalues $t$ are defined as
\begin{equation}\notag
\ket{\sigma,t,q^j;s,m}:=\sqrt{2\pi\hbar}\sum_{\sigma'}\op{P}_{\sigma',m}\ket{\sigma, t, q^j;s} \ ,
\end{equation}
and can also be written as follows:
\begin{equation}\label{eq:eigenstates-insertion-Dobs-t}
\begin{aligned}
&\left<\left.\frac{p_t}{c},\vec{p}\right|\sigma, t, q^j;s,m\right> = \frac{\left|p_1\right|^{\frac{1}{2}}}{(2\pi\hbar)^{\frac{d}{2}}}\Theta\left(\sigma p_1\right)\e^{-\frac{\I}{\hbar}cs p_1}\\
&\times\e^{-\frac{\I}{\hbar}t p_t}\e^{-\frac{\I}{\hbar}\sum_{j = 2}^d q^jp_j}\delta\left(-\frac{p_t^2}{2c^2}+\frac{\vec{p}^2}{2}+\frac{m^2c^2}{2}\right) \ .
\end{aligned}
\end{equation}
As before, these states evolve unitarily in $s$ and their evolution is generated by the Dirac observable $c\op{p}_1$:
\begin{equation}
\I\hbar\frac{\partial}{\partial s}\ket{\sigma, t, q^j;s,m} = c\op{p}_1\ket{\sigma, t, q^j;s,m} \ .
\end{equation}
The Dirac observable~[cf.~(\ref{eq:observable-practical}) and~(\ref{eq:observable-practical-1})]
\begin{equation}\label{eq:observable-practical-2}
\begin{aligned}
&\op{\mathcal{O}}_m[t|q^1 = cs]=\sum_{\sigma = \pm}\int_{-\infty}^{\infty}\D ct\\
&\times\int_{\mathbb{R}^{d-1}}\D^{d-1}q\ t\ket{\sigma,t, q^j;s,m}\bra{\sigma,t, q^j;s,m} \ 
\end{aligned}
\end{equation}
obeys the Heisenberg equation of motion
\begin{equation}
\I\hbar\frac{\partial}{\partial s}\op{\mathcal{O}}_m[t|q^1 = cs] = c\left[\op{p}_1, \op{\mathcal{O}}_m[t|q^1 = cs]\right]\ ,
\end{equation}
which is the quantum version of~(\ref{eq:PB-classical-2}).

\subsubsection{Nonrelativistic limit}
The restriction of the eigenstate $\ket{\sigma,t,q^j;s,m}$ given in~(\ref{eq:eigenstates-insertion-Dobs-t}) to a given frequency sector is obtained by acting on this state with the operator $\Theta\left(-\frac{\sigma\op{p}_t}{c}\right)$. Using~(\ref{eq:eigenstates-insertion-Dobs-t}), the result can be written in the nonrelativistic limit as follows.
\begin{align*}
&\left<ct,\vec{q}\left|\Theta\left(-\frac{\sigma\op{p}_t}{c}\right)\right|\tilde{\sigma}, \tilde{t},\tilde{q}^j;s,m\right>\\
&=\int\frac{\D^d p}{(2\pi\hbar)^{\frac{2d+1}{2}}}\ \frac{\e^{-\frac{\I}{\hbar}\sigma(t-\tilde{t})\sqrt{\vec{p}^2c^2+m^2c^4}}}{\sqrt{\vec{p}^2+m^2c^2}}\e^{\frac{\I}{\hbar}p_1(q^1-cs)}\\
&\times\e^{\frac{\I}{\hbar}\sum_{j = 2}^dp_j(q^j-\tilde{q}^j)}\Theta(\tilde{\sigma}p_1)|p_1|^{\frac{1}{2}}\\
&=\frac{\e^{-\frac{\I}{\hbar}\sigma mc^2(t-\tilde{t})}}{\sqrt{2\pi\hbar}\ mc}\int\frac{\D^d p}{(2\pi\hbar)^d}\ \e^{-\frac{\I}{\hbar}\sigma\frac{\vec{p}^2}{2m}(t-\tilde{t})}\e^{\frac{\I}{\hbar}p_1(q^1-cs)}\\
&\times\e^{\frac{\I}{\hbar}\sum_{j = 2}^dp_j(q^j-\tilde{q}^j)}\Theta(\tilde{\sigma}p_1)|p_1|^{\frac{1}{2}}+\mathcal{O}\left(\frac{1}{c^3}\right)\\
&= \sqrt{2\pi\hbar}\frac{\e^{-\frac{\I}{\hbar}\sigma mc^2(t-\tilde{t})}}{mc}\\
&\times\left<t,\vec{q}\left|\op{P}_{\sigma,m}^{\text{nonrel}}\Theta\left(\tilde{\sigma}\op{p}_1\right)|\op{p}_1|^{\frac{1}{2}}\right|\tilde{t}, q^1 = cs,\tilde{q}^j\right> +\mathcal{O}\left(\frac{1}{c^3}\right)\ ,
\end{align*}
where $\op{P}_{\sigma,m}^{\text{nonrel}}$ was defined in~(\ref{eq:nonrel-propagator-projector}). From the above equation we conclude that
\begin{equation}\notag
\begin{aligned}
&\left<ct',\vec{q}'\left|\Theta\left(-\frac{\sigma\op{p}_t}{c}\right)\op{\mathcal{O}}_m[t|q^1 = cs]\Theta\left(-\frac{\sigma\op{p}_t}{c}\right)\right|ct, \vec{q}\right>\\
&=\frac{\e^{-\frac{\I}{\hbar}\sigma mc^2(t'-t)}}{mc}\left<t',\vec{q}'\left|\op{\mathcal{O}}_{\sigma,m}^{\text{nonrel}}[t|q^1 = cs]\right|t, \vec{q}\right>\!+\! \mathcal{O}\left(\frac{1}{c^3}\right) ,
\end{aligned}
\end{equation}
where~[cf.~(\ref{eq:observable-practical-nonrel-1}) and~(\ref{eq:observable-practical-2})]
\begin{align}
&\op{\mathcal{O}}_{\sigma,m}^{\text{nonrel}}[t|q^1 = cs]:= 2\pi\hbar\sum_{\tilde{\sigma} = \pm}\int_{-\infty}^{\infty}\D \tilde{t}\notag\\
&\times\int_{\mathbb{R}^{d-1}}\D^{d-1}\tilde{q}\ \tilde{t}\ \op{P}_{\sigma,m}^{\text{nonrel}}\Theta\left(\tilde{\sigma}\op{p}_1\right)\left|\frac{\op{p}_1}{m}\right|^{\frac{1}{2}}\ket{\tilde{t}, q^1 = cs, \tilde{q}^j}\notag\\
&\times\bra{\tilde{t}, q^1 = cs, \tilde{q}^j}\left|\frac{\op{p}_1}{m}\right|^{\frac{1}{2}}\Theta\left(\tilde{\sigma}\op{p}_1\right)\op{P}_{\sigma,m}^{\text{nonrel}} \label{eq:observable-practical-nonrel-2}
\end{align}
is the nonrelativistic invariant extension of $\op{t}$. The matrix element of $\op{\mathcal{O}}_{\sigma,m}^{\text{nonrel}}[t|q^1 = cs]$ between two test functions $\psi^{(1,2)}(p_t,\vec{p})$ of compact support, which satisfy $\psi^{(1,2)}(p_t,p_1 = 0,p_j) = 0$, can be computed in analogy to the derivation of~(\ref{eq:insertion-Dobs-q}),~(\ref{eq:non-rel-insertion}), and~(\ref{eq:insertion-Dobs-t}). The result is
\begin{equation}\label{eq:non-rel-insertion-2}
\begin{aligned}
&\left<\psi^{(1)}\left|\op{\mathcal{O}}_{\sigma,m}^{\text{nonrel}}[t|q^1 = cs]\right|\psi^{(2)}\right>\\
&=\int\D p_t\D^dp\ \bar{\psi}^{(1)}(p_t,\vec{p})\delta\left(\sigma p_t+\frac{\vec{p}^2}{2m}\right)\\
&\times\left\{\I\hbar\frac{\partial}{\partial p_t}-\I\hbar\frac{\sigma m}{p_1}\frac{\partial}{\partial p_1}+\frac{\sigma m}{p_1}cs+\I\hbar\frac{\sigma m}{2p_1^2}\right\}\psi^{(2)}(p_t,\vec{p}) \ .
\end{aligned}
\end{equation}
Equation~(\ref{eq:non-rel-insertion-2}) corresponds to an operator insertion into the induced inner product of the two test functions. The inserted operator is symmetric in the auxiliary inner product and commutes with the Schr\"odinger constraint. It is a symmetric quantization of the classical Newtonian time-of-arrival observable given in~(\ref{eq:nonrel-DObs-2}) if the following identifications are made: $t(\tau)\to\I\hbar\frac{\partial}{\partial p_t}, q^1(\tau)\to\I\hbar\frac{\partial}{\partial p_1}, q^1(a) \to cs$.

It is worthwhile to compare the above result with previous works in the literature. In~\cite{Grot:1996,Hoehn:2018-2}, the time-of-arrival operator was carefully analyzed. In~\cite{Grot:1996}, it was concluded that a regularization of the operator was necessary in order to render it self-adjoint in the ``reduced'' Hilbert space associated with the degrees of freedom $\op{\vec{q}}, \op{\vec{p}}$ (there were no corresponding $p_t, \I\hbar\frac{\partial}{\partial p_t}$ operators). This regularization was then extended in~\cite{Hoehn:2018-2} to the complete operator inserted in~(\ref{eq:non-rel-insertion-2}), i.e. with the $p_t, \I\hbar\frac{\partial}{\partial p_t}$ terms included. The purpose of~\cite{Hoehn:2018-2} was to relate the ``Dirac quantization program'', which is the quantization based on building the physical Hilbert space from the kernel of the constraint operator and using the Rieffel induced inner product (cf.~Sec.~\ref{sec:general-quantum}), with the ``reduced phase-space quantization'', in which fewer variables are promoted to operators (in particular, the variable gauge-fixed to be time is not quantized). The relation between the two quantization strategies was established in~\cite{Hoehn:2018-2} for the simple model of the parametrized nonrelativistic particle and it relied on the notion of ``trivialization maps,''\footnote{Similar constructions were also analyzed in the literature in different contexts. See, for instance,~\cite{Creutz:1979} for an application to non-Abelian gauge fields and~\cite{Arlen-Anderson:1993} in the context of quantum canonical transformations.} which are isometries between the physical Hilbert space and the reduced Hilbert spaces. The view expressed in~\cite{Hoehn:2018-2} was that the relational content of a generally covariant quantum theory can be fully appreciated only if the Dirac and reduced quantization programs are used concomitantly and related via the trivialization maps.

We have taken a different attitude in the present article. Our point of view is that the Dirac quantization program is sufficient and captures all the relational content of the theory, provided one is equipped with a method of construction of invariant extensions of operators. We have proposed such a method~(in Sec.~\ref{sec:quantum-Dobs}) based on the usual Faddeev-Popov construction. The reason we have adopted this view is due to the fact that the relational dynamics is encapsulated in Dirac observables (evolving constants) already in the classical theory and the reduced phase space can be entirely understood from the construction of such observables~\cite{HT:book}. In this way, the Dirac quantization picture of a physical Hilbert space, on which the eigenstates of Dirac observables evolve unitarily, is enough (as we presented in Secs.~\ref{sec:rel-qDobs-1} and~\ref{sec:rel-qDobs-2} and as will be discussed in Sec.~\ref{sec:kasner}). Moreover, we have not used a regularization for the operator inserted in~(\ref{eq:non-rel-insertion-2}), as was done in~\cite{Grot:1996,Hoehn:2018-2}. The reason for this is that our Dirac observable is defined to be the operator given in~(\ref{eq:observable-practical-2}) in the relativistic case and in~(\ref{eq:observable-practical-nonrel-2}) in the nonrelativistic limit. The eigenstates of these operators form an orthonormal system in the induced inner product, which is sufficient for our purposes. These operators only coincide with the usual time-of-arrival operator given in~(\ref{eq:non-rel-insertion-2}), strictly speaking, for test functions of compact support\footnote{The generalized eigenstates of~(\ref{eq:observable-practical-nonrel-2}) or their relativistic counterparts given in~(\ref{eq:eigenstates-insertion-Dobs-t}) do not have compact support.} in the nonrelativistic limit. Nevertheless, it may be that a rigorous regularization procedure becomes necessary in more realistic applications of the method here described.

\section{\label{sec:kasner}The Kasner Model}
Let us now briefly illustrate how the method of construction of quantum observables here presented can be used in quantum cosmology. We consider the simplest anisotropic cosmology: the vacuum Bianchi I (Kasner) model. For a detailed discussion of this and other anisotropic cosmologies, see~\cite{Nick:2019,Ryan:book,Ellis:book} and references therein.

\subsection{Classical theory}
The Bianchi I model can be obtained by the symmetry-reduced ansatz for the spacetime metric
\begin{equation}
\D s^2 = -N^2\D\tau^2+a_x^2\D x^2+a_y^2\D y^2+a_z^2\D z^2 \ ,
\end{equation}
where $\tau$ is the time coordinate and $N$ is the lapse function. It is convenient to adopt the ``Misner variables'' $\alpha, \beta_+, \beta_-$, which can be defined as follows.
\begin{equation}
\begin{aligned}
a_x &= \e^{\alpha+\beta_+ +\sqrt{3}\beta_-} \ ,\\ a_y &= \e^{\alpha+\beta_+ - \sqrt{3}\beta_-} \ , \\ a_z &= \e^{\alpha-2\beta_+} \ .
\end{aligned}
\end{equation}
The scale factor of the universe is $\left(a_x a_y a_z\right)^{\frac{1}{3}} = e^{\alpha}$. The symmetry-reduced Einstein-Hilbert action reads\footnote{In this section, following~\cite{Nick:2019}, we adopt units in which $\frac{3c^6 V_0}{4\pi G} = 1$, where $V_0$ is the volume of space and $G$ is Newton's constant.}
\begin{equation}\label{eq:EH-bianchi}
S = \frac{1}{2}\int\D\tau\ \frac{e^{3\alpha}}{N}\left(-\dot{\alpha}^2+\dot{\beta}_+^2+\dot{\beta}_-^2\right) \ ,
\end{equation}
which is reparametrization invariant. The Misner variables are worldline scalars. After a Legendre transformation, Eq.~(\ref{eq:EH-bianchi}) leads to the Hamiltonian
\begin{equation}\label{eq:kasner-hamiltonian}
\mathcal{H} = \frac{N\e^{-3\alpha}}{2}\left(-p_{\alpha}^2+p_+^2+p_-^2\right) \ .
\end{equation}
The momentum conjugate to the lapse function is constrained to vanish. Thus, the lapse plays the role of an arbitrary multiplier in the Hamiltonian formulation and it can, without loss of generality, be chosen to be $N(\tau) = \e^{3\alpha(\tau)}e(\tau)$, where $e(\tau)$ is the einbein, such that~(\ref{eq:kasner-hamiltonian}) takes the form given in~(\ref{eq:primary-Hamiltonian}) with $C = -\frac{p_{\alpha}^2}{2}+\frac{p_+^2}{2}+\frac{p_-^2}{2}$. In this way, the vacuum Bianchi I model corresponds to a free massless relativistic particle~[cf.~(\ref{eq:rel-constraint})] in $2+1$ dimensions and the formalism presented in Sec.~\ref{sec:rel} can be applied in the limit $m\to0$. In particular, we will be interested in the invariant extension of the scale factor with respect to the gauge condition $\frac{\beta_+}{p_+}$. This can be conveniently written as~[cf.~(\ref{eq:general-relational-scalar-classical}) and~(\ref{eq:proper-time-evolution})]
\begin{equation}\label{eq:kasner-obs-integral}
\begin{aligned}
&\mathcal{O}[e^{\alpha}|\beta_+-p_+s = 0]\\
&:= \int_{-\infty}^{\infty}\D\eta\ \left|p_+\right|\delta\left(\beta_+(\eta)-p_+(\eta)s\right) e^{\alpha(\eta)} \ ,
\end{aligned}
\end{equation}
where $\eta = \int\D\tau\ e(\tau)$ is the proper time parameter. From~(\ref{eq:kasner-hamiltonian}), one readily finds
\begin{equation}
\begin{aligned}
\alpha(\eta) &= \alpha -p_{\alpha}\eta \ , \ p_{\alpha}(\eta) = p_{\alpha} \ ,\\
\beta_{\pm}(\eta) &= \beta_{\pm}+p_{\pm}\eta \ , \ p_{\pm}(\eta) = p_{\pm} \ .
\end{aligned}
\end{equation}
In this way, eq.~(\ref{eq:kasner-obs-integral}) can be explicitly computed. We obtain
\begin{equation}\label{eq:kasner-obs-explicit}
\mathcal{O}[e^{\alpha}|\beta_+-p_+s = 0] = \exp\left(\alpha+\frac{p_{\alpha}}{p_+}\beta_+-p_{\alpha}s\right) \ ,
\end{equation}
which can easily be seen to Poisson commute with the constraint. The scale factor vanishes (the singularity is reached) when $p_{\alpha}s\to\infty$. 

\subsection{Quantum theory}
Our goal is now to assess whether the classical singularity can be avoided in the quantum theory by analyzing the behavior of wave packets associated with the eigenstates of the invariant extension of the scale factor. We will consider that the singularity is avoided if the transition probability from the wave packet to the state which corresponds to the classical singularity is zero. This is a version of DeWitt's criterion~\cite{DeWitt:1967}, which is often employed in a heuristic manner without an associated inner product or probability interpretation (see, for instance,~\cite{Nick:2019}). Here, we are able to apply this criterion in a more complete fashion by assuming the Born rule remains valid for the transition probabilities computed with the induced inner product\footnote{See, however, the discussion in Sec.~\ref{sec:conclusions}.}.

As in Secs.~\ref{sec:rel-qDobs-1} and~\ref{sec:rel-qDobs-2}, we now construct the matrix elements of the invariant extension of the scale factor with respect to the gauge condition $\op{\chi}(s) = \op{\beta}_+-\op{p}_+s$, which is a self-adjoint operator in the auxiliary inner product. The eigenstates of the gauge condition satisfy
\begin{align*}
\op{\chi}(s)\ket{\chi, \alpha, \beta_{-};s} &= \chi\ket{\chi, \alpha, \beta_{-};s} \ ,
\end{align*}
and they can be written as
\begin{equation}\label{eq:eigenstates-chi-1}
\begin{aligned}
&\braket{p_{\alpha},p_+, p_-|\chi, \alpha, \beta_{-};s}\\
&= \frac{1}{(2\pi\hbar)^{\frac{3}{2}}}\e^{-\frac{\I}{\hbar}p_+\chi}\e^{-\frac{\I}{\hbar}\frac{p_+^2}{2}s}\e^{-\frac{\I}{\hbar}p_{\alpha}\alpha}\e^{-\frac{\I}{\hbar}p_{-}\beta_-} \ .
\end{aligned}
\end{equation}
It is straightforward to verify that the states given in~(\ref{eq:eigenstates-chi-1}) form an orthonormal system in the auxiliary inner product (for a fixed value of $s$). Since $\op{p}_+$ is a Dirac observable, the eigenstates of the Dirac observable associated with the scale factor are~[cf.~(\ref{eq:general-relational-scalar-quantum}) and~(\ref{eq:ket-omega-2})]
\begin{equation}\label{eq:kasner-eigenstate}
\begin{aligned}
&\ket{\sigma, \alpha,\beta_-;s} := \sqrt{2\pi\hbar}\sum_{\sigma' = \pm}\op{P}_{\sigma', m = 0}\left|\op{p}_+\right|^{\frac{1}{2}}\\
&\times\Theta(\sigma\op{p}_+)\ket{\chi = 0, \alpha, \beta_{-};s} \ ,
\end{aligned}
\end{equation}
where $\op{P}_{\sigma', m = 0}$ is the improper projector onto a given frequency sector of the massless relativistic particle~[cf.~(\ref{eq:matrix-P-momentum})]. The physical transition amplitude between two such states is found to be
\begin{align*}
&\left(\sigma', \alpha',\beta_-';s'|\sigma,\alpha,\beta_-;s\right)\\
&=\delta_{\sigma', \sigma}\int\frac{\D p_{\alpha}\D p_-\D p_+}{(2\pi\hbar)^2}\e^{\frac{\I}{\hbar}\frac{p_+^2}{2}(s'-s)}\e^{\frac{\I}{\hbar}p_{\alpha}(\alpha'-\alpha)}\\
&\times\e^{\frac{\I}{\hbar}p_{-}(\beta_-'-\beta_-)}\Theta(\sigma p_+)|p_+|\delta\left(-\frac{p_{\alpha}^2}{2}+\frac{p_+^2}{2}+\frac{p_-^2}{2}\right)\\
&=\delta_{\sigma', \sigma}\int\frac{\D p_{\alpha}\D p_-}{(2\pi\hbar)^2}\e^{\frac{\I(s'-s)}{2\hbar}\left(p_{\alpha}^2-p_-^2\right)}\e^{\frac{\I}{\hbar}p_{\alpha}(\alpha'-\alpha)}\e^{\frac{\I}{\hbar}p_{-}(\beta_-'-\beta_-)} \ .
\end{align*}
If $s'\to s$, this reduces to $\delta_{\sigma', \sigma}\delta(\alpha'-\alpha)\delta(\beta_-'-\beta_-)$. In general, we obtain~[cf.~(\ref{eq:nonrel-propagator-1})]
\begin{equation}\label{eq:gauge-fixed-propagator-1}
\begin{aligned}
&\left(\sigma', \alpha',\beta_-';s'|\sigma,\alpha,\beta_-;s\right)\\
&= \delta_{\sigma',\sigma}\overline{K}_{(\alpha)}(\alpha',s';\alpha,s)K_{(-)}(\beta_-',s';\beta_-,s) \ ,
\end{aligned}
\end{equation}
where $K_{(\alpha)}(\alpha',s';\alpha,s)$ and $K_{(-)}(\beta_-',s';\beta_-,s)$ are the usual (nonrelativistic) propagators
\begin{equation}\label{eq:gauge-fixed-propagator-2}
\begin{aligned}
&K_{(\alpha)}(\alpha',s';\alpha,s)\\
&= \left[2\pi\hbar\I(s'-s)\right]^{-\frac{1}{2}}\exp\left(-\frac{(\alpha'-\alpha)^2}{2\I\hbar(s'-s)}\right) \ , \\
&K_{(-)}(\beta_-',s';\beta_-,s)\\
&= \left[2\pi\hbar\I(s'-s)\right]^{-\frac{1}{2}}\exp\left(-\frac{(\beta_-'-\beta_-)^2}{2\I\hbar(s'-s)}\right)\ .
\end{aligned}
\end{equation}
We thus define~[cf.~(\ref{eq:general-relational-scalar-quantum}) and~(\ref{eq:kasner-obs-integral})]
\begin{equation}\label{eq:kasner-extensions}
\begin{aligned}
&\op{\mathcal{O}}[f(\alpha,\beta_-)|\chi(s) = 0]\\
&:= \sum_{\sigma = \pm}\int\D\alpha\D\beta_-\ f(\alpha,\beta_-)\ket{\sigma, \alpha,\beta_-;s}\bra{\sigma, \alpha,\beta_-;s} \ .
\end{aligned}
\end{equation}
The invariant extension of the scale factor is obtained by setting $f(\alpha,\beta_-) = e^{\alpha}$. For simplicity, let us consider the Gaussian wave packet
\begin{equation}\notag
\begin{aligned}
&\ket{\psi,\sigma;s}:= \int\D\alpha\D\beta_-\ \psi_{(\alpha)}(\alpha)\psi_{(-)}(\beta_-)\ket{\sigma, \alpha,\beta_-;s} \ ,\\
&\psi_{(\alpha)}(\alpha) := \left[\pi\mathcal{A}^2\right]^{-\frac{1}{4}}\e^{\frac{\I}{\hbar}p_{\alpha}^0(\alpha-\alpha_0)}\e^{-\frac{(\alpha-\alpha_0)^2}{2\mathcal{A}^2}} \ , \\
&\psi_{(-)}(\beta_-) := \left[\pi\mathcal{B}^2\right]^{-\frac{1}{4}}\e^{\frac{\I}{\hbar}p_{-}^0(\beta_--\beta_0)}\e^{-\frac{(\beta_--\beta_0)^2}{2\mathcal{B}^2}} \ ,
\end{aligned}
\end{equation}
which is normalized in the induced inner product, i.e., $(\psi,\sigma';s|\psi,\sigma;s) = \delta_{\sigma', \sigma}$. Using~(\ref{eq:gauge-fixed-propagator-1}) and~(\ref{eq:gauge-fixed-propagator-2}), we can compute the physical overlap
\begin{equation}\label{eq:kasner-physical-evolution}
\begin{aligned}
&(\sigma',\alpha',\beta_-'; s|\psi,\sigma; s= 0)\\
&= \delta_{\sigma', \sigma}\left[\int_{-\infty}^{\infty}\D\alpha\ \overline{K}_{(\alpha)}(\alpha',s;\alpha,0)\psi_{(\alpha)}(\alpha)\right]\\
&\times\left[\int_{-\infty}^{\infty}\D\beta_-\ K_{(-)}(\beta_-',s;\beta_-,0)\psi_{(-)}(\beta_-)\right]\\
&=:\delta_{\sigma',\sigma}\psi_{(\alpha)}(\alpha';s)\psi_{(-)}(\beta_-';s) \ ,
\end{aligned}
\end{equation}
where
\begin{align}
\psi_{(\alpha)}(\alpha;s) &:= \left[\pi^{\frac{1}{2}}\left(\mathcal{A}-\frac{\I\hbar s}{\mathcal{A}}\right)\right]^{-\frac{1}{2}}\e^{\frac{\I}{\hbar}p_{\alpha}^0\left(\alpha-\alpha_0+\frac{1}{2}p_{\alpha}^0s\right)}\notag\\
&\times\exp\left[-\frac{(\alpha-\alpha_0+p_{\alpha}^0s)^2}{2\mathcal{A}^2\left(1-\frac{\I\hbar s}{\mathcal{A}^2}\right)}\right]\ , \label{eq:kasner-psi-alpha-s}\\
\psi_{(-)}(\beta_-;s) &:= \left[\pi^{\frac{1}{2}}\left(\mathcal{B}+\frac{\I\hbar s}{\mathcal{B}}\right)\right]^{-\frac{1}{2}}\e^{\frac{\I}{\hbar}p_{-}^0\left(\beta_--\beta_0-\frac{1}{2}p_{-}^0s\right)}\notag\\
&\times\exp\left[-\frac{(\beta_--\beta_0-p_{-}^0s)^2}{2\mathcal{B}^2\left(1+\frac{\I\hbar s}{\mathcal{B}^2}\right)}\right]\ . \label{eq:kasner-psi-beta-minus-s}
\end{align}
The transition probability associated with~(\ref{eq:kasner-physical-evolution}) is
\begin{align*}
&\left|(\sigma',\alpha',\beta_-'; s|\psi,\sigma; s= 0)\right|^2\\
&= \delta_{\sigma',\sigma}\left|\psi_{(\alpha)}(\alpha';s)\right|^2\left|\psi_{(-)}(\beta_-';s)\right|^2 \ .
\end{align*}
Using~(\ref{eq:kasner-psi-alpha-s}) and~(\ref{eq:kasner-psi-beta-minus-s}), we obtain
\begin{equation}\label{eq:kasner-DeWitt-criterion}
\lim_{|\alpha'|\to\infty}\left|(\sigma',\alpha',\beta_-'; s|\psi,\sigma; s= 0)\right|^2 = 0 \ ,
\end{equation}
i.e., the probability for the transition from the Gaussian wave packet to the invariant eigenstate with a zero scale factor eigenvalue vanishes. We take this to mean the singularity is avoided (for Gaussian wave packets). Incidentally, if we define the uncertainty of an observable $\op{\mathcal{O}}$ as
\begin{equation}\label{eq:kasner-uncertainty}
\Delta\mathcal{O} = \left<\left(\op{\mathcal{O}}-\braket{\op{\mathcal{O}}}\right)^2\right>^{\frac{1}{2}}\ , 
\end{equation}
where $\braket{\cdot}$ denotes the average taken with respect to the induced inner product, then a simple calculation confirms that $\ket{\psi,\sigma;s=0}$ is a ``minimum uncertainty wave packet'' in the sense that the following equalities are satisfied~[cf.~(\ref{eq:kasner-extensions})]:
\begin{equation}
\begin{aligned}
&\Delta\mathcal{O}[\alpha|\chi(s=0) = 0]\Delta p_{\alpha} = \frac{\hbar}{2} \ ,\\
&\Delta\mathcal{O}[\beta_-|\chi(s=0) = 0]\Delta p_- = \frac{\hbar}{2} \ .
\end{aligned}
\end{equation}
Similarly, we can compute the expectation value~[cf.~(\ref{eq:kasner-extensions}) and~(\ref{eq:kasner-physical-evolution})]
\begin{align*}
&\left<\op{\mathcal{O}}[\e^{\alpha}|\chi(s) = 0]\right>\\
&=\sum_{\sigma' = \pm}\int\D\alpha\D\beta_-\ \e^{\alpha}\left|(\psi,\sigma,s=0|\sigma', \alpha,\beta_-;s)\right|^2\\
&=\int_{-\infty}^{\infty}\D\alpha\ \e^{\alpha}\ \left|\psi_{(\alpha)}(\alpha;s)\right|^2 \ .
\end{align*}
Using~(\ref{eq:kasner-psi-alpha-s}), we find
\begin{equation}\label{eq:kasner-quantum-obs-explicit}
\begin{aligned}
&\left<\op{\mathcal{O}}[\e^{\alpha}|\chi(s) = 0]\right>\\
&= \exp\left[\alpha_0-p_{\alpha}^0s+\frac{1}{4}\left(\mathcal{A}^2+\frac{\hbar^2s^2}{\mathcal{A}^2}\right)\right] \ .
\end{aligned}
\end{equation}
This expression is to be compared with its classical counterpart~(\ref{eq:kasner-obs-explicit}). Notably, the expectation value given in~(\ref{eq:kasner-quantum-obs-explicit}) does not vanish for any value of $s$, in contrast to~(\ref{eq:kasner-obs-explicit}) in the classical theory. In fact, Eq.~(\ref{eq:kasner-quantum-obs-explicit}) describes a quantum bounce. The minimum value of the average scale factor,
\begin{equation}
\begin{aligned}
&\left<\op{\mathcal{O}}[\e^{\alpha}|\chi(s) = 0]\right>_{\text{min}}\\
&= \exp\left[\alpha_0-\frac{(p_{\alpha}^0)^2\mathcal{A}^2}{\hbar^2}+\frac{\mathcal{A}^2}{4}\right]\ ,
\end{aligned}
\end{equation}
is reached when
\begin{equation}
s = s_{\text{bounce}} = \frac{2p_{\alpha}^0\mathcal{A}^2}{\hbar^2} \ .
\end{equation}
Besides (the probabilistic version of) DeWitt's criterion~(\ref{eq:kasner-DeWitt-criterion}), Eq.~(\ref{eq:kasner-quantum-obs-explicit}) is another indication that the classical singularity may be avoided in the quantum theory, at least for the minimum uncertainty wave packet. This illustrates how the method of construction of Dirac observables presented in Sec.~\ref{sec:quantum-Dobs} may be used to obtain concrete results in quantum cosmology regarding the dynamics of gauge-invariant operators and their associated invariant transition amplitudes.

\vspace{-0.35cm}
\section{\label{sec:conclusions}Conclusions}
\vspace{-0.15cm}
Although there is currently no consensus about the correct way to quantize the gravitational field, promising approaches can be developed if the gauge symmetry of the theory is genuinely understood. Following~\cite{Rovelli:1990-1,Rovelli:1990-2,Rovelli:1991,Hartle:1996,Gambini:2001-1,Gambini:2001-2,Gambini:2009,Rovelli:2011,Tambornino:2012,Rovelli:2014,Hoehn:2018-1,Hoehn:2018-2,Hoehn:2019}, we have taken the view that the dynamical content of a generally covariant theory is relational and can be comprehended through gauge-invariant extensions of the dynamical variables. Classically, such extensions can be constructed by the Faddev-Popov procedure, which expresses gauge-fixed variables in an arbitrary gauge by means of integral formulas. In this article, we have indicated how this is realized in a model-independent way in terms of worldline diffeomorphisms in the case of generally covariant classical mechanics.

We have then translated this construction into the canonical (operator-based) quantum theory and we described how invariant extensions of operators and their eigenstates can be systematically constructed. We believe such a method was currently lacking in the literature and we have compared it to previous proposals. In particular, we stressed that our method differs from the one used in~\cite{Marolf:1995-1}, in which the invariant extension of the identity operator was not the identity. In contrast, we take this to be the defining property of the construction here presented. Evidently, the method we present will possibly need to be refined or made more rigorous in more realistic applications (e.g., in generally covariant canonical field theories).

Our method was exemplified for the case of the free relativistic particle and the related vacuum Bianchi I model. We have shown in detail how different quantum Dirac observables can be assembled and emphasized that their eigenstates evolve unitarily with respect to the (arbitrary) gauge-fixed time variable. Thus, there is no problem of time for the evolution of such observables. The dynamics is understood in the same way as in the classical theory and it depends on the choice of time parameter. In particular, the vacuum Bianchi I example demonstrates the usefulness of the method for concrete applications in quantum cosmology. Currently, we are working on the application of this construction of Dirac observables to more realistic cosmologies and we will report on this topic in the near future.

We have also analyzed the connection between the relational view adopted in this paper with another popular approach to the problem of time: the semiclassical emergence of WKB time~\cite{Kuchar:1991,Isham:1992,Kiefer:1993-1,Kiefer:1993-2,Zeh:1987,Kiefer:1993-1,Kiefer:1993-2}. This approach is relevant because many important phenomenological applications of quantum cosmology have been developed in this semiclassical framework~\cite{Halliwell:1984,Kiefer:2011-1,Brizuela:2015-1,Brizuela:2015-2,Kamenshchik:2013,Kamenshchik:2014,Kamenshchik:2016,Kamenshchik:2017}. Thus, it is worthwhile to understand its relation to the more complete quantum theory based on the physical Hilbert space and associated quantum Dirac observables. Indeed, the semiclassical approach to the problem of time invites a series of questions regarding its foundations, such as whether there is a Hilbert space for the full wave function(al) of gravitational and matter fields, or whether the dynamics is unitary beyond the semiclassical level.

In the present article, we have answered these questions. It is possible to construct a physical Hilbert space for all dynamical fields, which is based on the Rieffel induced inner product~\cite{Rieffel:1974,Landsman:1995,Marolf:1995-4,Marolf:1997,Marolf:2000,Halliwell:2001,Halliwell:2002} and on which quantum Dirac observables act as operators and evolve unitarily beyond the semiclassical regime. This is, in our view, the fundamental picture. We have then shown that the emergence of WKB time occurs in the weak-coupling (here, nonrelativistic) limit of invariant transition amplitudes defined with respect to the induced inner product. While this result is expected, it had not been shown before. This completes the discussion of~\cite{Chataig:2019}, in which it was argued that all the results of the semiclassical approach coincide with a choice of gauge (time variable) and can be extended beyond the semiclassical level, and it answers the question raised in~\cite{Kamen:2019} about the connection between the WKB time approach and gauge fixing methods. The results here presented suggest that the phenomenological work of~\cite{Kiefer:2011-1,Brizuela:2015-1,Brizuela:2015-2} concerning quantum gravitational corrections to the Cosmic Microwave Background power spectrum can be reinterpreted as the weak-coupling limit of a more fundamental theory based on the induced inner product. It is an intriguing open question whether these corrections can be refined using the construction of Dirac observables we have presented and we leave this topic for future work.

Finally, it is worthwhile to clarify that although we take the view that the physical Hilbert space based on the induced inner product is the correct and more fundamental space on which the relational dynamics of quantum Dirac observables can be defined, it is far from clear whether the Born rule should still be valid or modified in this context. We have tacitly used the Born rule throughout and also explicitly to discuss singularity avoidance in the Kasner model. While there is no problem of time in the sense that the (unitary) evolution of gauge-invariant operators can be defined with respect to different choices of the time parameter, there is still the measurement problem, which becomes even more vexed in a generally covariant theory. The formalism here presented remains silent on this issue and we hope to return to this in the future.

\begin{acknowledgments}
The author thanks Claus Kiefer for useful discussions, and the Bonn-Cologne Graduate School of Physics and Astronomy for financial support.
\end{acknowledgments}


\end{document}